\documentclass[iop]{emulateapj}
\usepackage{subfigure}
\usepackage{apjfonts}
\usepackage{amsmath}
\usepackage{amssymb}
\usepackage{gensymb}
\usepackage{color}

\def \Se#1{Section~\ref{sec:#1}}
\def \Ses#1#2{Sections~\ref{sec:#1}-\ref{sec:#2}}

\def \Fig#1{Fig.~\ref{fig:#1}}

\def \Figure#1{Figure~\ref{fig:#1}}

\def \Tbl#1{Table~\ref{tbl:#1}}

\def \Eqn#1{Equation~\ref{eqn:#1}}

\def \u{{$u^*$}}
\def \g{{$g^{\prime}$}}
\def \r{{$r^{\prime}$}}
\def \i{{$i^{\prime}$}}
\def \z{{$z^{\prime}$}}

\def \Mg{{$M_{g^{\prime}}$~}}

\def \Reg{{$R_{e,g^{\prime}}$}}

\def \zz{{$z$~}}
\def \MS{{$M_*$~}}
\def \Msun{{M$_{\odot}$}}
\def \arcsec{{$^{\prime\prime}$}}

\def \eg{{e.g.}}
\def \ie{{i.e.~}}
\def \rms{{\it rms~}}
\def \etal{{et al.}}

\slugcomment{For submission to ApJ.}

\shorttitle{NGVS XXIV. CMR and Comparisons with Models}
\shortauthors{Roediger et al. 2016}

\begin{document}


\title{The Next Generation Virgo Cluster Survey (NGVS). XXIV. The Red Sequence to $\sim$10$^6$ L$_{\odot}$\\ and Comparisons with Galaxy Formation Models.}

\author{Joel C. Roediger$^{1}$, Laura Ferrarese$^{1}$, Patrick C{\^o}t{\'e}$^{1}$, Lauren A. MacArthur$^{1,2}$, R{\'u}ben S{\'a}nchez-Janssen$^{1}$, John P. Blakeslee$^{1}$, Eric W. Peng$^{3,4}$, Chengze Liu$^{5}$, Roberto Munoz$^{6}$, Jean-Charles Cuillandre$^{7}$, Stephen Gwyn$^{1}$, Simona Mei$^{8,9}$, Samuel Boissier$^{10}$, Alessandro Boselli$^{10}$, Michele Cantiello$^{11,12}$, St{\'e}phane Courteau$^{13}$, Pierre-Alain Duc$^{7}$, Ariane Lan\c{c}on$^{14}$, J. Christopher Mihos$^{15}$, Thomas H. Puzia$^{6}$, James E. Taylor$^{16}$, Patrick R. Durrell$^{17}$, Elisa Toloba$^{18,19}$, Puragra Guhathakurta$^{18}$, Hongxin Zhang$^{6,20}$}

\affil{$^1$ National Research Council of Canada, Herzberg Astronomy and Astrophysics Program, 5071 West Saanich Road, Victoria, BC, V9E 2E7, Canada}
\affil{$^2$ Department of Astrophysical Sciences, Princeton University, Princeton, NJ 08544, USA}
\affil{$^3$ Department of Astronomy, Peking University, Beijing 100871, China}
\affil{$^4$ Kavli Institute for Astronomy and Astrophysics, Peking University, Beijing 100871, China}
\affil{$^5$ Center for Astronomy and Astrophysics, Department of Physics and Astronomy, Shanghai Jiao Tong University, Shanghai 200240, China}
\affil{$^6$ Instituto de Astrof{\'i}sica, Pontificia Universidad Cat{\'o}lica de Chile, Av. Vicu{\~n}a Mackenna 4860, 7820436 Macul, Santiago, Chile}
\affil{$^7$ AIM Paris Saclay, CNRS/INSU, CEA/Irfu, Universit{\'e} Paris Diderot, Orme des Merisiers, 91191 Gif sur Yvette Cedex, France}
\affil{$^8$ GEPI, Observatoire de Paris, CNRS, 5 Place Jules Jannssen - 92195 Meudon, France}
\affil{$^9$ Universit{\'e} Paris Denis Diderot, 75205, Paris Cedex 13, France}
\affil{$^{10}$ Aix Marseille Universit{\'e}, CNRS, Laboratoire d'Astrophysique de Marseille UMR 7326, 13388, Marseille, France}
\affil{$^{11}$ INAF Osservatorio Astr. di Teramo, via Maggini,-64100, Teramo, Italy}
\affil{$^{12}$ INAF Osservatorio Astr. di Capodimonte, Salita Moiariello, 16, 80131 Napoli, Italy}
\affil{$^{13}$ Department of Physics, Engineering Physics and Astronomy, Queen's University, Kingston, ON, Canada}
\affil{$^{14}$ Observatoire Astronomique, Universit{\'e} de Strasbourg \& CNRS UMR 7550, 11 rue de l'Universit{\'e}, 67000 Strasbourg, France}
\affil{$^{15}$ Department of Astronomy, Case Western Reserve University, Cleveland, OH, USA}
\affil{$^{16}$ Department of Physics and Astronomy, University of Waterloo, Waterloo, ON N2L 3G1, Canada}
\affil{$^{17}$ Department of Physics and Astronomy, Youngstown State University, Youngstown, OH, USA}
\affil{$^{18}$ UCO/Lick Observatory, University of California, Santa Cruz, 1156 High Street, Santa Cruz, CA 95064, USA}
\affil{$^{19}$ Department of Physics, Texas Tech University, Box 41051, Lubbock, TX, USA}
\affil{$^{20}$ National Astronomical Observatories, Chinese Academy of Sciences, Beijing 100012, China}

\email{Joel.Roediger@nrc-cnrc.gc.ca}


\begin{abstract}
We use deep optical photometry from the Next Generation Virgo Cluster Survey 
[NGVS] to investigate the color-magnitude diagram for the galaxies inhabiting 
the core of this cluster.  The sensitivity of the NGVS imaging allows us to 
continuously probe galaxy colors over a factor of $\sim 2 \times 10^5$ in 
luminosity, from brightest cluster galaxies to scales overlapping classical 
satellites of the Milky Way [\Mg $\sim$ --9; \MS $\sim 10^6$ \Msun], within a 
single environment.  Remarkably, we find the first evidence that the RS flattens 
in all colors at the faint-magnitude end [starting between --14 $\le$ \Mg $\le$ 
--13, around \MS $\sim 4 \times 10^7$ \Msun], with the slope decreasing to 
$\sim$60\% or less of its value at brighter magnitudes.  This could indicate 
that the stellar populations of faint dwarfs in Virgo's core share similar 
characteristics [\eg~constant mean age] over $\sim$3 mags in luminosity, 
suggesting that these galaxies were quenched coevally, likely via pre-processing 
in smaller hosts.  We also compare our results to galaxy formation models, 
finding that the RS in model clusters have slopes at intermediate magnitudes 
that are too shallow, and in the case of semi-analytic models, do not reproduce 
the flattening seen at both extremes [bright/faint] of the Virgo RS.  
Deficiencies in the chemical evolution of model galaxies likely contribute to 
the model-data discrepancies at all masses, while overly efficient quenching may 
also be a factor at dwarf scales.  Deep UV and near-IR photometry are required 
to unambiguously diagnose the cause of the faint-end flattening.
\end{abstract}



\section{Introduction}\label{sec:intro}

Despite the complexities of structure formation in a $\Lambda$CDM Universe, 
galaxies are well-regulated systems.  Strong evidence supporting this statement 
are the many fundamental relations to which galaxies adhere: star formation rate 
versus stellar mass or gas density \citep{Daddi07, Elbaz07, Noeske07, KE12}, 
rotational velocity versus luminosity or baryonic mass for disks 
\citep{Courteau07, McGaugh12, Lelli16}, the fundamental plane for spheroids 
\citep{Bernardi03, Zaritsky12}, and the mass of a central compact object versus 
galaxy mass \citep{Ferrarese06, WH06, Beifiori12, KH13}, to name several.  
Moreover, many of these relations are preserved within galaxy groups and 
clusters, demonstrating that such regulation is maintained in all environments 
\citep[\eg][]{BM09}.  This paper focusses on the relationship between color and 
luminosity for quiescent [``quenched''] galaxies: the so-called red sequence 
[RS].

First identified by \cite{dV61} and \cite{VS77}, the RS represents one side of 
the broader phenomenon of galaxy color bimodality \citep{Strateva01, Blanton03, Baldry04, Balogh04, Driver06, Cassata08, Taylor15}, the other half being the 
blue cloud, with the green valley separating them.  Based on the idea of 
passively evolving stellar populations, color bimodality is widely interpreted 
as an evolutionary sequence where galaxies transform their cold gas into stars 
within the blue cloud and move to the RS after star formation ends \citep[\eg][]{Faber07}.  This evolution has been partly observed through the increase of mass 
density along the RS towards low redshift \citep{Bell04, Kriek08, Pozzetti10}, 
although the underlying physics of quenching remains a matter of active 
research.  The standard view of color bimodality is a bit simplistic though 
insofar as the evolution does not strictly proceed in one direction; a fraction 
of galaxies in the RS or green valley have their stellar populations temporarily 
rejuvenated by replenishment of their cold gas reservoirs \citep{Schawinski14}.

Crucial to our understanding of the RS is knowing when and how it formed.  The 
downsizing phenomenon uncovered by spectroscopic analyses of nearby early-type 
galaxies \citep[ETGs;][]{Nelan05, Thomas05, Choi14} implies that the RS was 
built over an extended period of time [$\sim$5 Gyr], beginning with the most 
massive systems \citep[\eg][]{Tanaka05}.  These results support the common 
interpretation that the slope of the RS is caused by a decline in the 
metallicity [foremost] and age of the constituent stellar populations towards 
lower galaxy masses \citep{KA97, Ferreras99, Terlevich99, Poggianti01, DL07}.  
Efforts to directly detect the formation of the RS have observed color 
bimodality to \zz $\sim$ 2 \citep{Bell04, Willmer06, Cassata08}.  More recently, 
legacy surveys such as GOODS, COSMOS, NEWFIRM, and UltraVISTA have shown that 
massive quiescent galaxies [\MS $\gtrsim 3 \times 10^{10}$ \Msun] begin to 
appear as early as \zz = 4 \citep{Fontana09, Muzzin13, Marchesini14} and finish 
assembling by \zz = 1-2 \citep{Ilbert10, Brammer11}.  Growth in the stellar mass 
density of quiescent galaxies since \zz = 1, on the other hand, has occured at 
mass scales of $M^*$ and lower \citep{Faber07}, consistent with downsizing.

Owing to their richness, concentration, and uniform member distances, galaxy 
clusters are an advantageous environment for studying the RS.  Moreover, their 
characteristically high densities likely promote quenching and therefore hasten 
the transition of galaxies to the RS.  In terms of formation, the RS has been 
identified in [proto-]clusters up to \zz $\sim$ 2 \citep{Muzzin09, Wilson09, Gobat11, Spitler12, Stanford12, Strazzullo13, Cerulo16}.  Much of the interest 
in \zz $>$ 0 clusters has focussed on the growth of the faint end of the RS.  
Whereas scant evidence has been found for evolution of either the slope or 
scatter of the RS \citep[but see \citealt{Hao09} and \citealt{Hilton09}]{Ellis97, Gladders98, Stanford98, Blakeslee03, Holden04, Lidman08, Mei09, Papovich10}, 
several groups have claimed an elevated ratio of bright-to-faint RS galaxies in 
clusters up to \zz = 0.8, relative to local measurements \citep[see also \citealt{BG14} and references therein]{Smail98, DL07, Stott07, Gilbank08, Hilton09, Rudnick09}.  The increase in this ratio with redshift indicates that 
low-mass galaxies populate the RS at later times than high-mass systems, meaning 
that the former, on average, take longer to quench and/or are depleted via 
mergers/stripping at early epochs.  These results are not without controversy, 
however, with some arguing that the inferred evolution may be the result of 
selection bias, small samples, or not enough redshift baseline 
\citep{Crawford09, Lu09, DP13, Andreon14, Romeo15, Cerulo16}.

As a tracer of star formation activity and stellar populations, colors also are 
a key metric for testing galaxy formation models.  Until recently, only 
semi-analytic models [SAMs] had sufficient statistitcs to enable meaningful 
comparisons to data from large surveys.  Initial efforts indicated that the 
fraction of red galaxies was too high in models, and thus quenching too 
efficient, which led to suggestions that re-accretion of SNe ejecta was 
necessary to maintain star formation in massive galaxies \citep{Bower06}.  Since 
then, a persistent issue facing SAMs has been that their RSs are shallower than 
observed \citep{Menci08, Gonzalez09, Guo11}.  The common explanation for this is 
that the stellar metallicity-luminosity relation in the models is likewise too 
shallow.  \cite{Font08} demonstrated that an added cause of the excessively red 
colors of dwarf satellites is their being too easily quenched by strangulation, 
referring to the stripping of halo gas.  While \cite{Font08} increased the 
binding energy of this gas as a remedy, \cite{GP14} have shown that further 
improvements are still needed.  Studies of other models have revealed similar 
mismatches with observations \citep{Romeo08, Weinmann11}, indicating that the 
problem is widespread.

In this paper, we use multi-band photometry from the Next Generation Virgo 
Cluster Survey \citep[NGVS;][]{Ferrarese12} to study galaxy colors in the core 
of a \zz = 0 cluster, an environment naturally weighted to the RS.  The main 
novelty of this work is that NGVS photometry probes mass scales from brightest 
cluster galaxies to Milky Way satellites \citep[hereafter F16]{Ferrarese16b}, 
allowing us to characterize the RS over an unprecedented factor of $>$10$^5$ in 
luminosity [$\sim$10$^6$ \Msun~in stellar mass] and thus reach a largely 
unexplored part of the color-magnitude distribution [CMD].  Given the unique 
nature of our sample, we also take the opportunity to compare our data to galaxy 
formation models, which have received scant attention in the context of cluster 
cores.

Our work complements other NGVS studies of the galaxy population within Virgo's 
core.  \cite{Zhu14} jointly modelled the dynamics of field stars and globular 
clusters [GCs] to measure the total mass distribution of M87 to a projected 
radius of 180 kpc.  \cite{Grossauer15} combined dark matter simulations and the 
stellar mass function to extend the stellar-to-halo mass relation down to $M_h 
\sim 10^{10}$ \Msun.  \cite{SJ16} statistically inferred the intrinsic shapes of 
the faint dwarf population and compared the results to those for Local Group 
dwarfs and simulations of tidal stripping.  \cite{Ferrarese16a} present the 
deepest luminosity function to date for a rich, volume-limited sample of nearby 
galaxies.  Lastly, C{\^o}t{\'e} \etal~(in preparation) and S{\'a}nchez-Janssen 
\etal~(in preparation) study the galaxy and nuclear scaling relations, 
respectively, for the same sample.

In \Se{data} we briefly discuss our dataset and preparations thereof.  Our 
analysis of the RS is presented in \Se{rs}, while \Ses{comp-prev}{comp-model} 
focus on comparisons to previous work, compact stellar systems [CSS] and galaxy 
formation models.  A discussion of our findings and conclusions are provided in 
\Ses{disc}{conc}.


\section{Data}\label{sec:data}

Our study of the RS in the core of Virgo is enabled by the NGVS 
\citep{Ferrarese12}.  Briefly, the NGVS is an optical imaging survey of the 
Virgo cluster performed with CFHT/MegaCam.  Imaging was obtained in the 
\u\g\i\z~bands\footnote{Note that the filters used in the NGVS are not identical 
to those of the Sloan Digital Sky Survey \citep[SDSS;][]{York00}, with the 
\u--band being the most different.  Unless otherwise stated, magnitudes are 
expressed in the MegaCam system throughout this paper.} over a 104 deg$^2$ 
footprint centered on sub-clusters A and B, reaching out to their respective 
virial radii (1.55 and 0.96 Mpc, respectively, for an assumed distance of 16.5 
Mpc; \citealt{Mei07}; \citealt{Blakeslee09}).  The NGVS also obtained \r--band 
imaging for an area of 3.71 deg$^2$ [0.3 Mpc$^2$], roughly centered on M87, the 
galaxy at the dynamical center of sub-cluster A; we refer to this as the core 
region.  NGVS images have a uniform limiting surface brightness of $\sim$29 
\g--mag arcsec$^{-2}$.  Further details on the acquisition and reduction 
strategies for the NGVS are provided in \cite{Ferrarese12}.

This paper focuses on the core of the cluster, whose boundaries are defined as,
\begin{center}
 $\begin{array}{rcrcl}
  12^{h}26^{m}20^{s} & \leq & \text{RA } ({\rm J2000}) & \leq & 12^{h}34^{m}10^{s} \nonumber\\
  11$\degree$30^{\prime}22$\arcsec$ & \leq & \text{Dec } ({\rm J2000}) & \leq & 13$\degree$26^{\prime}45$\arcsec$ \nonumber\\
 \end{array}$
\end{center}
and encompass four MegaCam pointings [see Figure 13 of F16].  A catalog of 404 
galaxies for this area, of which 154 are new detections, is published in F16, 
spanning the range 8.9 $\le$ \g~$\le$ 23.7 and $\ge$50\% complete to \g~$\sim$ 
22.  As demonstrated there, the galaxy photometry has been thoroughly analysed 
and cluster membership extensively vetted for this region; below we provide a 
basic summary of these endeavors.  A study of the CMD covering the entire survey 
volume will be presented in a future contribution.

Faint [\g~$>$ 16] extended sources in the core were identified using a dedicated 
pipeline based on ring-filtering of the MegaCam stacks.  Ring-filtering replaces 
pixels contaminated by bright, point-like sources with the median of pixels 
located just beyond the seeing disk.  This algorithm helps overcome situations 
of low surface brightness sources being segmented into several parts due to 
contamination.  The list of candidates is then culled and assigned membership 
probabilities by analysing SExtractor and GALFIT \citep{Peng02} parameters in 
the context of a size versus surface brightness diagram, colors and structural 
scaling relations, and photometric redshifts.  A final visual inspection of the 
candidates and the stacks themselves is made to address issues of 
false-positives, pipeline failures, and missed detections.  After this, the 
remaining candidates are assigned a membership flag indicating their status as 
either certain, likely, or possible members.

\begin{figure*}
 \begin{center}
  \subfigure[]{\includegraphics[trim=0 10 0 0, width=0.48\textwidth]{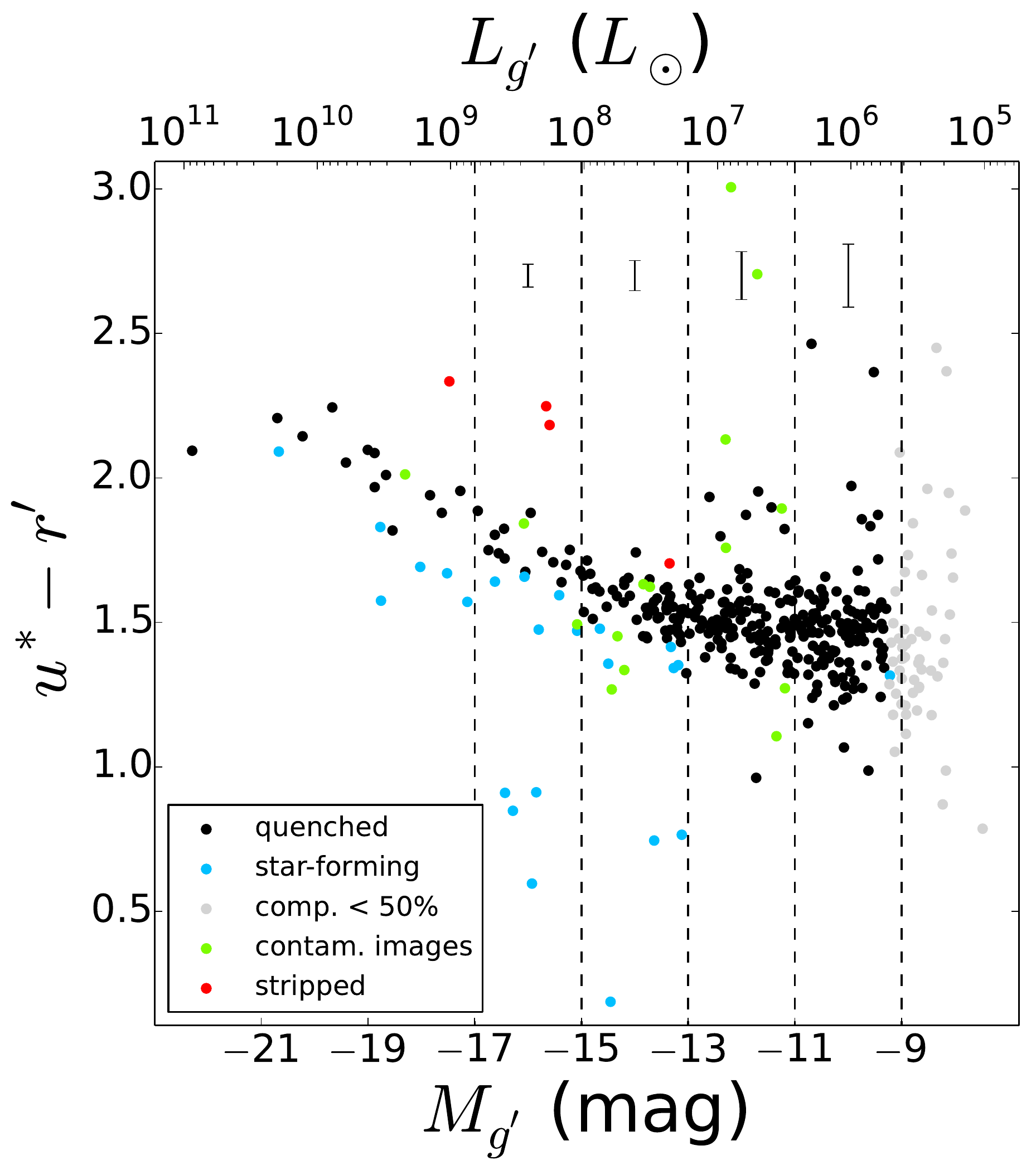}}
  \subfigure[]{\includegraphics[trim=0 10 0 0, width=0.48\textwidth]{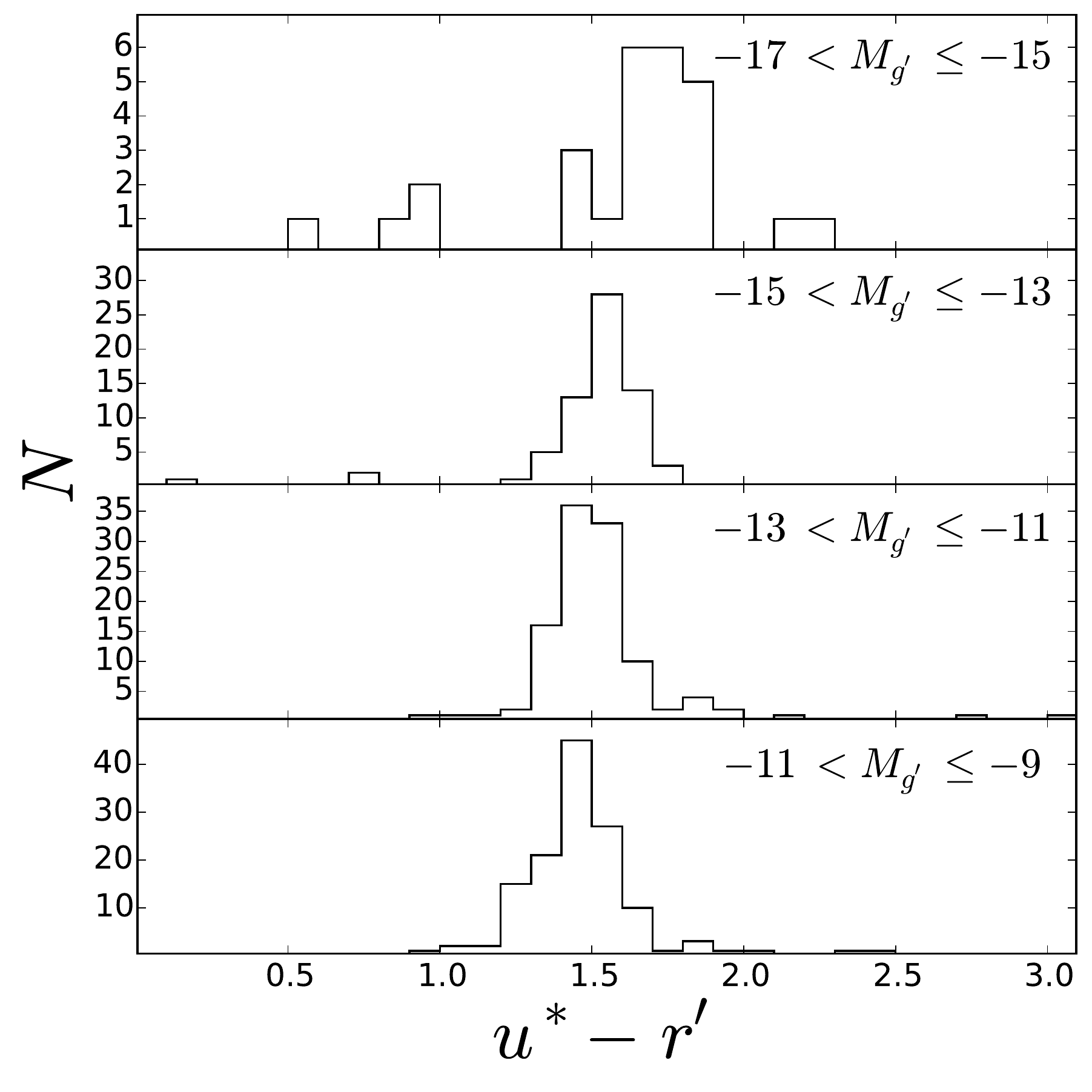}}
  \caption{({\it a}) (\u--\r) color versus absolute \g--band magnitude for the 
  404 galaxies in the core of Virgo.  Colored points are purged from our sample 
  of RS candidates due to obvious star formation activity [blue], our 
  completeness limits [grey], significant image contamination [green], or 
  suspected tidal stripping [red].  The vertical lines indicate bins of 
  magnitude referenced in the right-hand panel, with representative errors 
  plotted in each.  ({\it b}) Color distributions within the four magnitude bins 
  marked at left.  The NGVS photometry enables a deep study of the galaxy CMD 
  and we verify that the core of Virgo is highly deficient in star-forming 
  galaxies.}
  \label{fig:rs_demo}
 \end{center}
\end{figure*}

\begin{figure*}
 \begin{center}
  \includegraphics[width=1.0\textwidth]{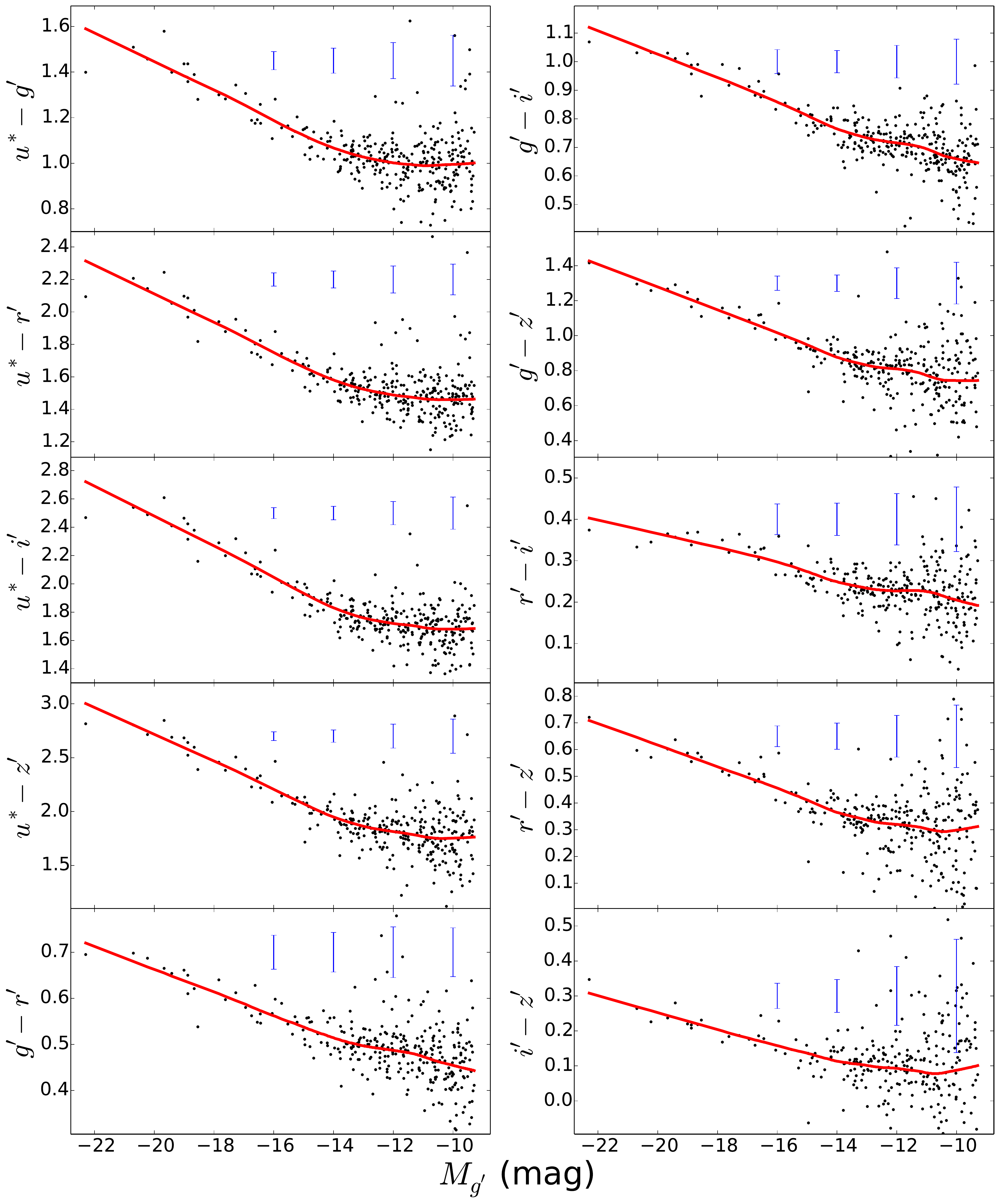}
  \caption{CMDs for quiescent galaxies in Virgo's core, in all ten colors 
  measured by the NGVS.  Fluxes have been measured consistently in all five 
  bands within apertures corresponding to 1.0 \Reg~isophote of each galaxy.  
  Black points represent individual galaxies while red lines show non-parametric 
  fits to the data.  The RS defines a color-magnitude relation in all colors 
  that flattens at faint magnitudes, which could be explained by a {\it 
  constant} mean age and metallicity for the lowest-mass galaxies in this region 
  [albeit with significant scatter; but see \Fig{comp-err}].  Representative 
  errors for the same magnitude bins as in \Fig{rs_demo} are shown in each 
  panel.}
  \label{fig:rs}
 \end{center}
\end{figure*}

As part of their photometric analysis, F16 measured surface brightness profiles 
and non-parametric structural quantities in the \u\g\r\i\z~bands for the core 
galaxies with the IRAF task ELLIPSE.  These data products are complemented with 
similar metrics from S{\'e}rsic fits to both the light profiles and image 
cutouts for each source [the latter achieved with GALFIT].  Our work is based on 
the growth curves deduced by applying their [non-parametric] \g--band isophotal 
solutions to all other bands while using a common master mask.  This allows us 
to investigate changes in the RS as a function of galactocentric radius, rather 
than rely on a single aperture.  \cite{Driver06} adopted a similar approach for 
their CMD analysis, finding that bimodality was more pronounced using core 
versus global colors; our results support this point [see \Fig{aper_eff}].  We 
extract from the growth curves all ten colors covered by the NGVS, integrated 
within elliptical apertures having semi-major axes of $a~\times$ \Reg~[\Reg~= 
\g--band effective radius], where $a$ = {0.5, 1.0, 2.0, 3.0}; we also examine 
colors corresponding to the total light of these galaxies.  Errors are 
estimated following \cite{Chen10}, using the magnitude differences between F16's 
growth curve and S{\'e}rsic analyses, and scaling values for each set of 
apertures by the fraction of light enclosed.  These estimates should probably be 
regarded as lower limits, since they do not capture {\it all} sources of 
systematic uncertainty.

Absolute magnitudes are computed assuming a uniform distance of 16.5 Mpc 
\citep{Mei07, Blakeslee09} for all galaxies and corrected for reddening using 
the York Extinction Solver\footnote{http://www4.cadc-ccda.hia-iha.nrc-cnrc.gc.ca/community/YorkExtinctionSolver/} \citep{McCall04}, adopting the \cite{Schlegel98} dust maps, \cite{Fitzpatrick99} extinction law, and $R_V$ = 3.07.  
To help gauge the intrinsic scatter along the RS, we use recovered magnitudes 
for $\sim$40k artificial galaxies injected into the image stacks [F16] to 
establish statistical errors in our total light measurements.  A more focussed 
discussion of uncertainties in the NGVS galaxy photometry may be found in 
\cite{Ferrarese16a} and F16.

We note that, although the NGVS is well-suited for their detection, 
ultra-compact dwarfs [UCDs] are excluded from our galaxy sample for two 
reasons.  First, they have largely been omitted from previous analyses of the 
RS.  Second, the nature of these objects is unsettled.  While many are likely 
the remnants of tidally-stripped galaxies \citep[\eg][]{Bekki03, Drinkwater03, PB13, Seth14}, the contribution of large GCs to this population remains 
unclear.  Readers interested in the photometric properties of the UCD population 
uncovered by the NGVS are referred to \cite{Liu15} for those found in the core 
region; still, we include UCDs in our comparisions of the colors of RS galaxies 
and CSS in \Se{comp-css}.


\section{The Red Sequence in the Core of Virgo}\label{sec:rs}

\Figure{rs_demo}a plots the (\u--\r) colors, integrated within 1.0 \Reg, of {\it 
all} 404 galaxies in the core of Virgo as a function of their total \g--band 
magnitudes.  One of the most striking features in this plot is the depth to 
which we probe galaxy colors: at its 50\%-completeness limit [\Mg $\sim$ --9], 
the NGVS luminosity function reaches levels that have only been previously 
achieved in the Local Group [\ie comparable to the Carina dSph, and only 
slightly brighter than Draco; \citealt{Ferrarese16a}].  This is significant as 
integrated colors for dwarf galaxies at these scales have, until now, been 
highly biased to the local volume [$D \le$ 4 Mpc], incomplete, and noisy 
\citep[\eg][]{Johnson13}.  The NGVS CMD therefore represents the most extensive 
one to date based on homogeneous photometry, spanning a factor of $2 \times 
10^5$ in luminosity.

Also interesting about \Fig{rs_demo}a is the dearth of blue galaxies in the core 
of Virgo.  This is more apparent in \Figure{rs_demo}b, where we plot histograms 
of (\u--\r) in four bins of luminosity.  Three of the four samples are well 
described as unimodal populations rather than the bimodal color distributions 
typically found in large galaxy surveys \citep[\eg][]{Baldry04}.  The absence of 
a strong color bimodality in Virgo's core is not surprising though 
\citep{Balogh04, Boselli14} and suggests that most of these galaxies have been 
cluster members long enough to be quenched by the environment\footnote{The 
timescale associated with environmental quenching appears contentious, with some 
groups favoring shorter values ($<$2 Gyr; \citealt{BG14}, and references 
therein; \citealt{Haines15}) and others longer \citep[several Gyr; \eg][]{Balogh00, DL12, Taranu14}.  Galaxy mass and possible delay times likely factor 
into this disagreement.}.  The minority of blue galaxies we find may be members 
that are currently making their first pericentric passage or are non-core 
members projected along the line-of-sight.  Since our interest lies in the RS, 
we have inspected three-color images for every galaxy and exclude from further 
analysis 24 that are clearly star-forming [blue points in \Fig{rs_demo}a].  
Also excluded are the 56 galaxies that fall below our completeness limit [grey 
points], 16 whose imaging suffers from significant contamination [\eg~scattered 
light from bright stars; green points], and 4 that are candidate remnants of 
tidal stripping [red points].  While we cannot rule out a contribution by 
reddening to the colors of the remaining galaxies, their three-color images do 
not indicate a significant frequency of dust lanes.

\begin{figure*}
 \begin{center}
  \includegraphics[width=1.0\textwidth]{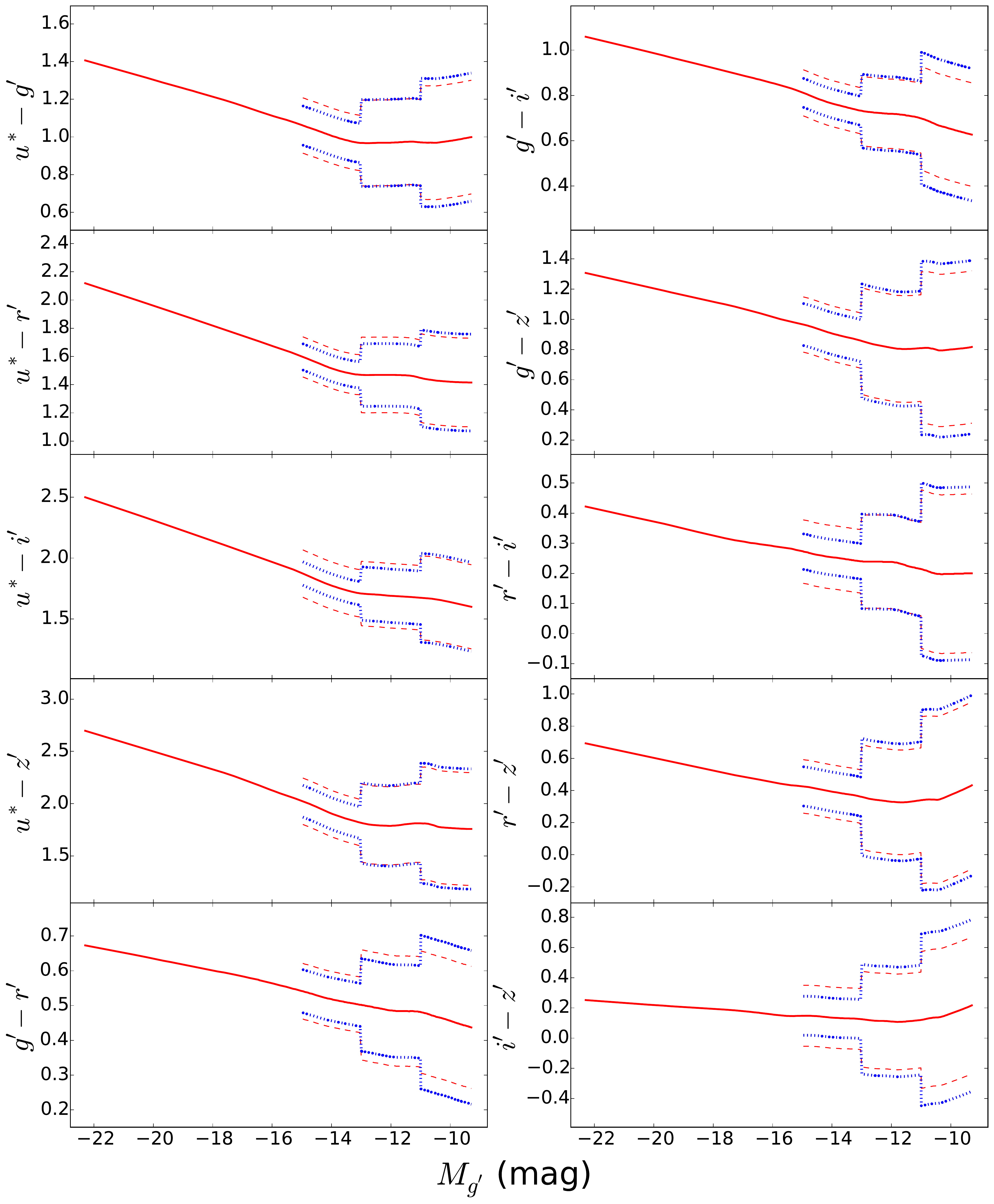}
  \caption{Comparison of the observed scatter [dashed lines] about the RS [solid 
  lines] to photometric errors [dotted lines] established from artificial galaxy 
  tests.  The comparision is limited to \Mg $\gtrsim$ --15 since our tests did 
  not probe brighter magnitudes.  The scatter and errors, averaged within three 
  bins of luminosity, match quite well, especially at the faintest luminosities, 
  suggesting minimal intrinsic scatter in the colors and stellar populations of 
  these galaxies.}
  \label{fig:comp-err}
 \end{center}
\end{figure*}

\Figure{rs} plots all ten CMDs for quiescent galaxy candidates in Virgo's core, 
where the colors again correspond to 1.0 \Reg.  Having culled the star-forming 
galaxies, we can straightforwardly study the shape of the RS as a function of 
wavelength.  In each panel of \Fig{rs} we observe a clear trend, whereby for \Mg 
$\lesssim$ --14, colors become bluer towards fainter magnitudes.  To help trace 
this, we have run the Locally Weighted Scatterplot Smoothing algorithm 
\citep[LOWESS;][]{Cleveland79} on each CMD; these fits are represented by the 
red lines in the figure.  The observed trends are notable given that optical 
colors are marginally sensitive to the metallicities of composite stellar 
populations with $Z \gtrsim 0.1 Z_{\odot}$.  Simple comparisons of our LOWESS 
curves to stellar population models suggests that, for \Mg $\lesssim$ --14, 
metallicity increases with luminosity along the RS [see \Fig{comp-k13}]; age 
trends are harder to discern with the colors available to us.  A 
metallicity-luminosity relation for RS galaxies agrees with previous work on the 
stellar populations of ETGs throughout Virgo \citep{Roediger11b} and the 
quiescent galaxy population at large \citep[\eg][]{Choi14}.  Our suggestion 
though is based on fairly restrictive assumptions about the star formation 
histories of these galaxies [\ie exponentially-declining, starting $\ge$8 Gyr 
ago]; more robust results on age and metallicity variations along the RS in 
Virgo's core from a joint UV-optical-NIR analysis will be the subject of future 
work.

A flattening at the bright end of the RS for Virgo ETGs was first identified by 
\cite{Ferrarese06} and later confirmed in several colors by 
\citet[hereafter JL09]{JL09}.  This seems to be a ubiquitous feature of the 
quiescent galaxy population, based on CMD analyses for nearby galaxies 
\citep{Baldry04, Driver06}.  This flattening may also be present in our data, 
beginning at \Mg $\sim$ --19, but the small number of bright galaxies in the 
core makes it difficult to tell.  Also, this feature does not appear in colors 
involving the \z--band, but this could be explained by a plausible error in 
this measurement for M87 [\eg~0.1 mag], the brightest galaxy in our sample.

The flattening seen at bright magnitudes implies that the RS is non-linear.  A 
key insight revealed by the LOWESS fits in \Fig{rs} is that the linearity of the 
RS also breaks down at faint magnitudes, in all colors.  The sense of this 
non-linearity is that, for \Mg $\gtrsim$ --14, the local slope is shallower than 
at brighter magnitudes, even flat in some cases [\eg~\u--\g; see Appendix].  For 
several colors [\eg~\r--\i], the LOWESS fits suggest that the behavior at the 
faint end of the RS may be even more complex, but the scale of such variations 
is well below the photometric errors [see \Fig{comp-err}].  JL09 found that 
the color-magnitude relation [CMR] of ETGs also changes slope in a similar 
manner, but at a brighter magnitude than us [\Mg $\sim$ --16.5]; we address this 
discrepancy in \Se{comp-prev}.

An implication of the faint-end flattening of the RS is that the low-mass dwarfs 
in Virgo's core tend to be more alike in color than galaxies of higher mass.  
This raises the question of whether the scatter at the faint-end of the RS 
reflects intrinsic color variations or just observational errors.  We address 
this issue in \Figure{comp-err} by comparing the observed scatter in the {\it 
total} colors to error estimates based on the artificial galaxies mentioned in 
\Se{data}.  Shown there are LOWESS fits to the data and the \rms scatter about 
them [solid and dashed lines, respectively], and the scatter expected from 
photometric errors [dotted lines].  Both types of scatter have been averaged 
within three bins of magnitude: --15 $<$ \Mg $\le$ --13, --13 $<$ \Mg $\le$ 
--11, and --11 $<$ \Mg $\le$ --9; the comparison does not probe higher 
luminosities because our artificial galaxy catalog was limited to \g~$>$ 16, by 
design.  We generally find that the scatter and errors both increase towards 
faint magnitudes and that the two quantities match well, except in the brightest 
bin, where the scatter mildly exceeds the errors.  For the other bins however, 
the intrinsic scatter must be small, strengthening the assertion that the 
faintest galaxies possess uniform colors [to within $\lesssim$0.05 mag] and, 
possibly, stellar populations.  Deeper imaging will be needed to improve the 
constraints on genuine color variations at these luminosities.

\begin{figure}
 \includegraphics[width=0.48\textwidth]{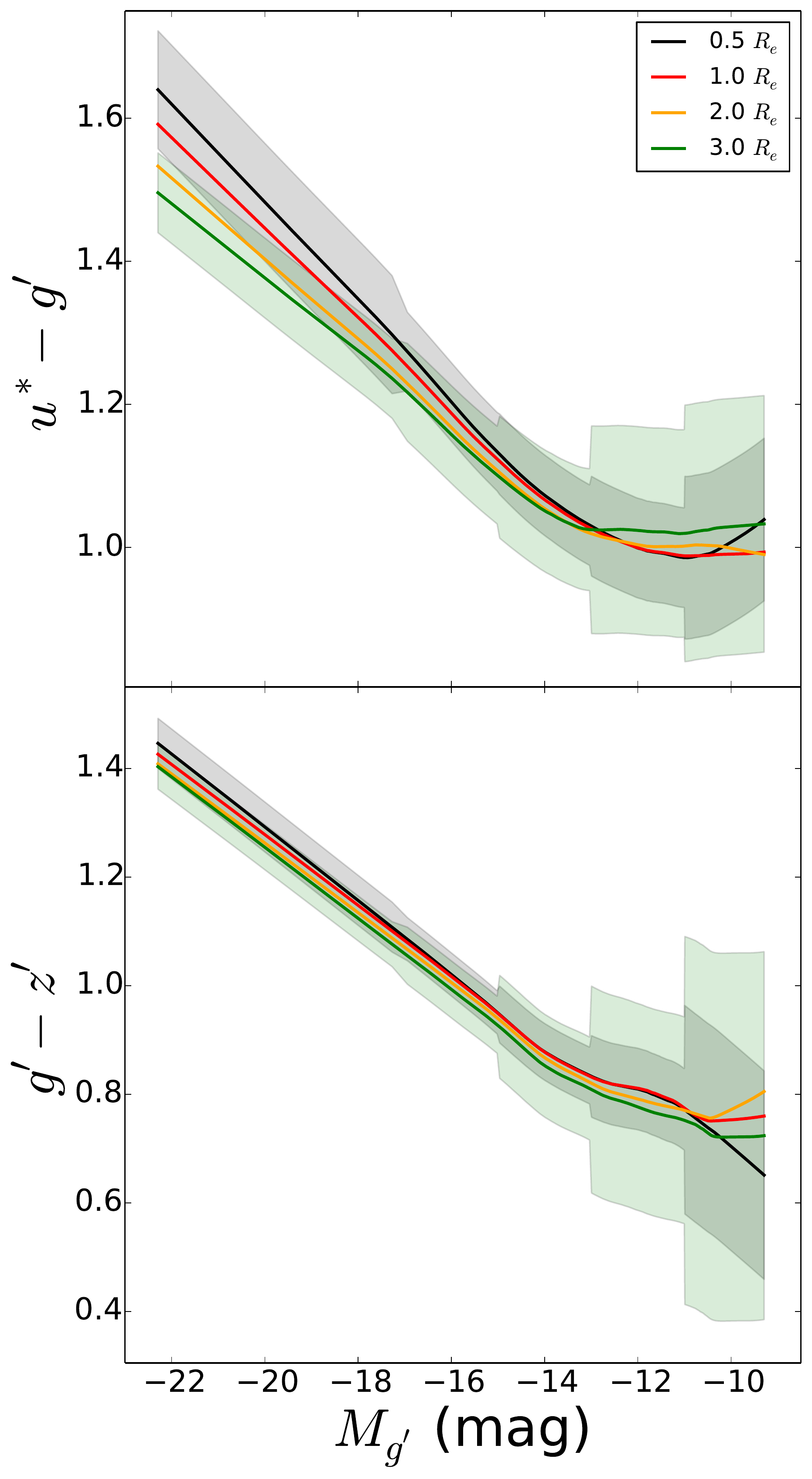}
 \caption{RS in (\u--\g) and (\g--\z), for different sizes of aperture used to 
 measure galaxy colors.  All four curves consider the same sample of galaxies.  
 The choice of aperture has an impact on the slope of the RS at \Mg $\lesssim$ 
 --16 mag for (\u--\g), with smaller apertures yielding steeper slopes, while 
 the RS is more stable in (\g--\z).  The shaded envelopes represent the scatter 
 about the RS for the 0.5 \Reg~and 3.0 \Reg~apertures.}
 \label{fig:aper_eff}
\end{figure}

The last topic examined in this section is the effect of aperture size on galaxy 
color.  Our most important result, the flattening of the RS at faint magnitudes, 
is based on galaxy colors integrated within their half-light radii.  Aperture 
effects could be significant in the presence of radial color gradients, as 
suggested by \cite{Driver06}, and therefore bias our inferences on the shape of 
the RS.  In \Figure{aper_eff} we show LOWESS fits to the \u--\g~and \g--\z~RSs 
for colors measured within 0.5 \Reg, 1.0 \Reg, 2.0 \Reg, and 3.0 \Reg.  These 
particular colors are chosen because, in the absence of deep UV and NIR 
photometry\footnote{UV and deep NIR imaging of the Virgo cluster exist 
\citep{Boselli11, Munoz14} but can only aid us for brighter galaxies and select 
fields, respectively.}, they provide the only leverage on stellar populations 
for the {\it full} NGVS dataset.  We also include measurements of the scatter 
about these fits for the 0.5 \Reg~and 3.0 \Reg~apertures, represented by the 
shaded envelopes.

The top panel of \Fig{aper_eff} shows that \u--\g~changes by at most 0.04-0.06 
mag at \Mg $\le$ --17 between consecutive aperture pairs.  Two-sample $t-$tests 
of linear fits to the data indicate that these differences are significant at 
the $P$ = 0.01 level.  Conversely, hardly any variation is seen between 
apertures for galaxies with \Mg > --16.  The bottom panel of \Fig{aper_eff} 
demonstrates that \g--\z~changes little with radius in most of our galaxies.  
Slight exceptions are the 0.5 \Reg~colors for galaxies with \Mg $\le$ --16, 
which differ from the 2.0 and 3.0 \Reg~colors by $\lesssim$0.04 mag.  The 1.0 
\Reg~colors bridge this gap, following the 0.5 \Reg~sequence at \Mg $\gtrsim$ 
--17 and moving towards the other sequences for brighter magnitudes.

The changes in the RS with galactocentric radius imply the existence of {\it 
negative} color gradients within specific regions of select galaxies.  The 
strongest gradients are found for \u--\g~within bright galaxies, inside 2.0 
\Reg, while galaxies with \Mg $>$ --15 have little-to-none in either color.  
Mild negative gradients are seen in \g--\z~between 0.5 and 1.0 \Reg~for galaxies 
with \Mg $<$ --17, consistent with previous work on the spatially-resolved 
colors of galaxies throughout Virgo \citep{Roediger11a}.  The most important 
insight though from \Fig{aper_eff} is that the flattening of the RS at faint 
magnitudes does not apply to a specific aperture.  The implications of those 
gradients we do detect in our galaxies, in terms of stellar populations and 
comparisons with galaxy formation models, will be addressed in \Se{disc}.


\section{Comparison to Previous Work}\label{sec:comp-prev}

Before discussing the implications of our results, over the next two sections we 
compare our RS to earlier/ongoing work on the colors of Virgo galaxies and CSS, 
starting with the former.  Of the several studies of the galaxy CMD in Virgo 
\citep{Bower92, Ferrarese06, Chen10, Roediger11a, Kim10}, that of JL09 is the 
most appropriate for our purposes.  JL09 measured colors for 468 ETGs from the 
Virgo Cluster Catalog \citep{Binggeli85}, based on $ugriz$ imaging from SDSS DR5 
\citep{AMC07}.  Their sample is spread throughout the cluster and has $B<$ 
18.0.  Most interestingly, they showed that these galaxies trace a non-linear 
relation in all optical CMDs, not unlike what we find for faint members 
inhabiting the centralmost regions.

\begin{figure}
 \includegraphics[width=0.48\textwidth]{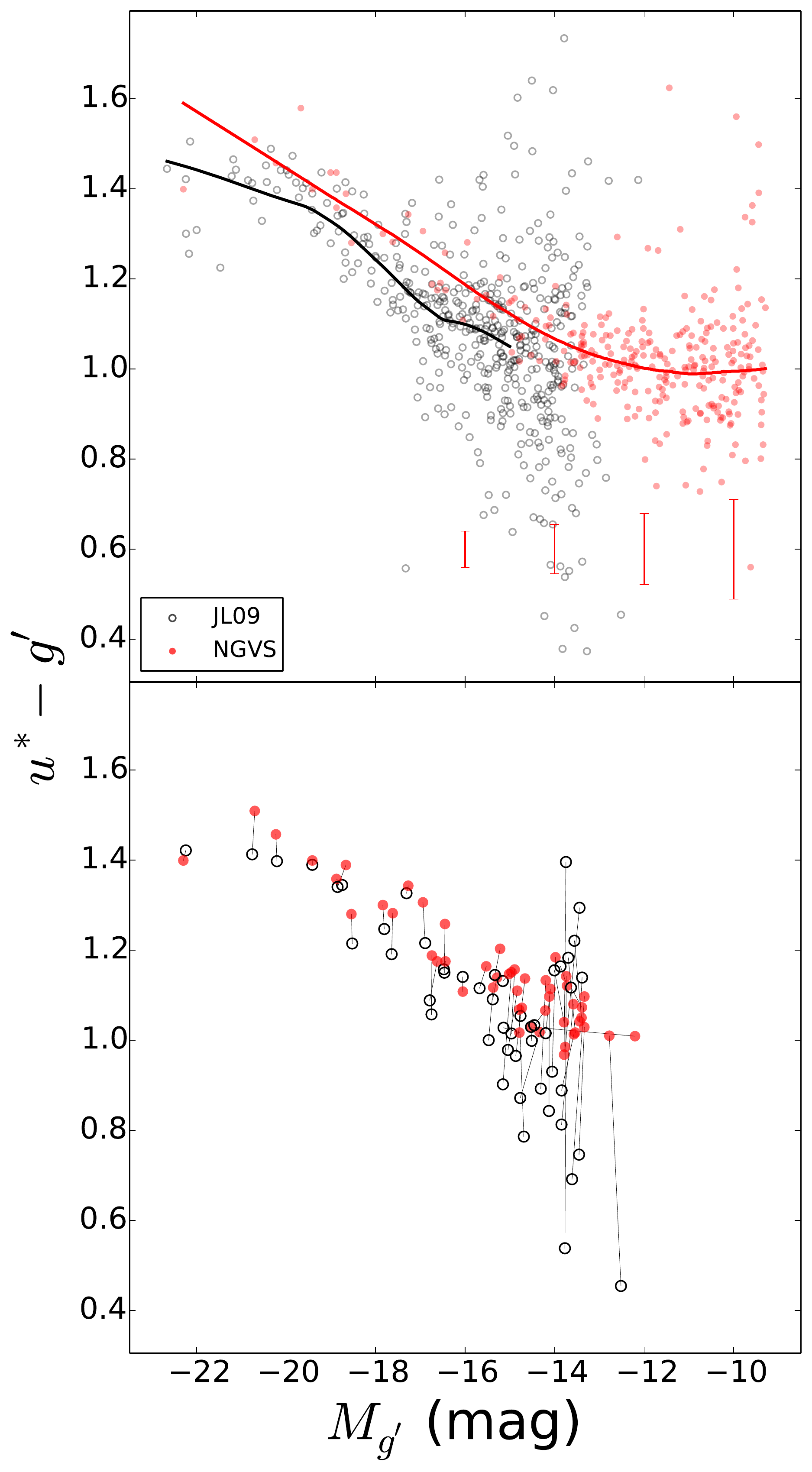}
 \caption{({\it top}) Comparison of the \u--\g~CMD from JL09 for Virgo ETGs 
 [black circles] to that measured here [red dots].  The full sample is plotted 
 for each dataset and LOWESS fits for both are overlaid [solid lines].  
 Representative errors for the NGVS are included along the bottom.  ({\it 
 bottom}) As above but restricted to the galaxies common to both samples; 
 measurement pairs are joined with lines.  The NGVS extends the CMD for this 
 cluster faintward by $\sim$5 mag, with much improved photometric errors.  We 
 also find that JL09's CMR is steeper than our own at intermediate magnitudes, 
 likely due to their inclusion of systems having recent star formation and 
 possible errors in their sky subtraction.}
 \label{fig:ngvs-vs-jl09}
\end{figure}

In \Figure{ngvs-vs-jl09} we overlay the \u--\g~CMD from JL09 against our own, 
measured within 1.0 \Reg; the comparison is appropriate since JL09 measured 
colors within their galaxies' $r$--band half-light radii.  We have transformed 
JL09's photometry to the CFHT/MegaCam system following Equation 4 in 
\cite{Ferrarese12}.  The top panel shows all objects from both samples, along 
with respective LOWESS fits, while the bottom is restricted to the 62 galaxies 
in common to both.  We focus on the \u--\g~color because of its importance to 
stellar population analyses; indeed, this is a reason why accurate \u--band 
photometry was a high priority for the NGVS.

The most notable feature in the top panel of \Fig{ngvs-vs-jl09} is the 
superior depth of the NGVS relative to the SDSS, an extension of $\sim$5 mag.  
There is a clear difference in scatter between the two samples, with that for 
JL09 increasing rapidly for \Mg $>$ --18, whereas the increase for the NGVS 
occurs much more gradually\footnote{While the scatter in the JL09 data is likely 
dominated by the shallower depth of the SDSS imaging, a contribution by distance 
uncertainties cannot be ruled out, since the Virgo Cluster Catalog spans several 
sub-groups whose relative distances can exceed 10 Mpc \citep{Mei07}.} [{\it cf}. 
\Fig{comp-err}; see Fig. 1 of \citealt{Ferrarese16a} as well].  Furthermore, the 
JL09 CMR has a lower zeropoint [by $\sim$0.06] and a shallower slope than the 
NGVS RS for \Mg $\lesssim$ --19, which two-sample $t-$tests verify as 
significant [$P$ = 0.01].  The JL09 data also exhibit a flattening of the CMR in 
the dwarf regime, but at a brighter magnitude than that seen in ours [\Mg $\sim$ 
--16.5].  The shallower slopes found by JL09 at both ends of their CMR are seen 
for other colors and so cannot be explained by limitations/biases in the SDSS 
$u$--band imaging.  The shallower slope at bright magnitudes substantiates what 
was hinted at in \Fig{rs} and is more obvious in JL09 since their sample covers 
the full cluster\footnote{Virgo comprises two large sub-clusters and several 
massive groups, such that its bright galaxies are spread throughout the 
cluster.}; the existence of this feature is also well-known from SDSS studies of 
the wider galaxy population \citep[\eg][]{Baldry04}.  The lower zeropoint of the 
JL09 CMR is seen in other colors too, hinting that calibration differences 
between SDSS DR5 and DR7 are responsible, where the NGVS is anchored to the 
latter \citep{Ferrarese12}.

Lastly, the LOWESS fits in \Fig{ngvs-vs-jl09} indicate that, between --19 
$\lesssim$ \Mg $\lesssim$ --16.5, the JL09 CMR has a steeper slope than the NGVS 
RS.  This difference is significant [$P$ = 0.01] and holds for other \u--band 
colors as well.  This steeper slope forms part of JL09's claim that the ETG CMR 
flattens at \Mg $\gtrsim$ --16.5, a feature not seen in our data.  Since JL09 
selected their sample based on morphology, recent star formation in dwarf 
galaxies could help create their steeper slope.  For one, the colors of many 
galaxies in the JL09 sample overlap with those flagged in our sample as 
star-forming.  Also, \cite{Kim10} find that dS0s in Virgo follow a steeper UV 
CMR than dEs and have bluer UV-optical colors at a given magnitude.  We 
therefore are unsurprised to have not observed the flattening detected by JL09.

Recent star formation cannot solely explain why JL09 find a steeper slope at 
intermediate magnitudes though.  The bottom panel of \Fig{ngvs-vs-jl09} shows 
that, for the same galaxies, JL09 measure systematically bluer \u--\g~colors; 
moreover, this difference grows to fainter magnitudes, creating a steeper CMR.  
Comparisons of other colors [\eg~\g--\r] and the agreement found therein proves 
that this issue only concerns JL09's $u$--band magnitudes.  The stated trend in 
the color discrepancy appears inconsistent with possible errors in our 
SDSS-MegaCam transformations.  Aperture effects can also be ruled out since the 
differences in size scatter about zero and never exceed 25\% for any one object; 
besides, \Fig{aper_eff} demonstrates that color gradients in \u--\g~are minimal 
at faint magnitudes.  A possible culprit may be under-subtracted backgrounds in 
JL09's $u$--band images since they performed their own sky subtraction.  
Therefore, we suggest that the differences between the JL09 CMR and NGVS RS for 
\Mg $>$ --19 can be explained by: (i) a drop in the red fraction amongst Virgo 
ETGs between --19 $\lesssim$ \Mg $\lesssim$ --16.5, and (ii) JL09's measurement 
of systematically brighter $u$--band magnitudes.  Despite this disagreement, 
these comparisons highlight two exciting aspects about the NGVS RS [and the 
photometry overall]: (i) it extends several magnitudes deeper than the SDSS, and 
(ii) the photometric errors are well-controlled up to the survey limits.


\section{Comparison to Compact Stellar Systems}\label{sec:comp-css}

\begin{figure*}
 \begin{center}
  \begin{tabular}{c}
   \includegraphics[width=1.0\textwidth]{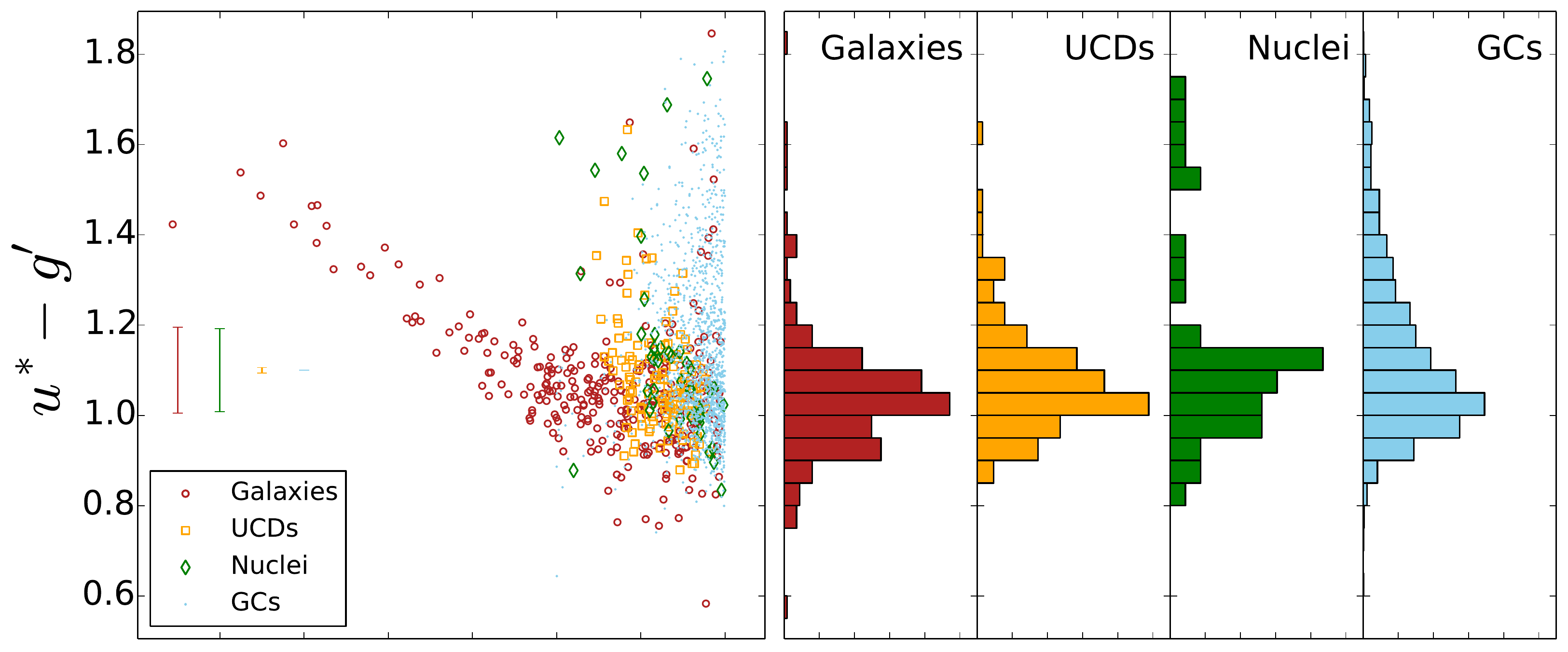}\cr
   \includegraphics[trim=0 10 0 0, width=1.0\textwidth]{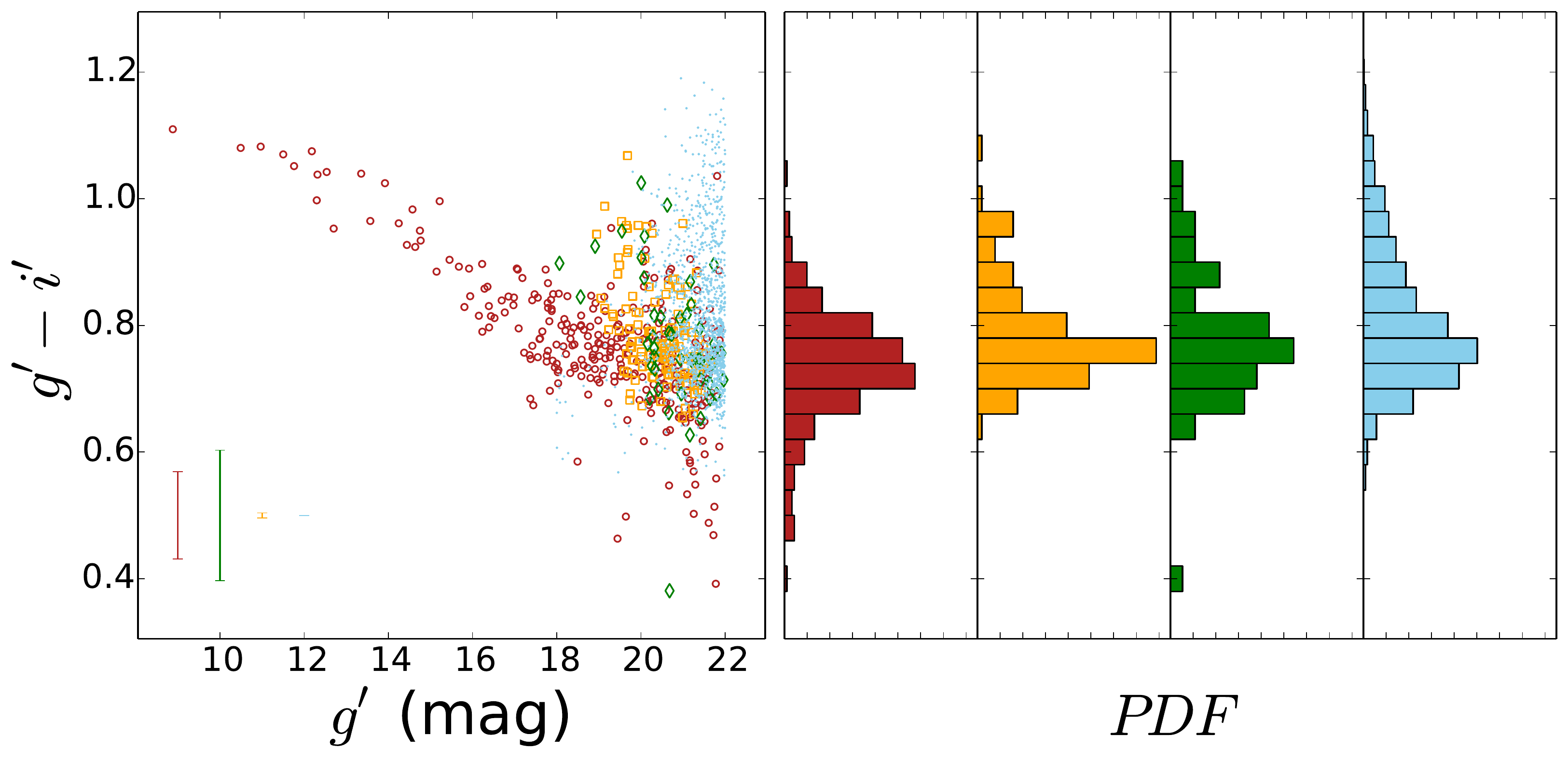}
  \end{tabular}
  \caption{({\it top row}) \u--\g~CMD and color distributions for galaxies 
  [circles], GCs [dots], UCDs [squares], and galactic nuclei [diamonds] within 
  the core of Virgo.  Since our intent is to compare these stellar systems 
  within a common magnitude range, only those CSS having 18 $<$ \g~$<$ 22 
  are plotted.  Representative errors for each population at faint magnitudes 
  are included at bottom-left.  ({\it bottom row}) As above but for the 
  \g--\i~color.  At faint magnitudes, comparitively red objects are only found 
  amongst the CSS populations; their colors are likely caused by a higher metal 
  content than those for galaxies of the same luminosity.}
  \label{fig:gal-vs-css}
 \end{center}
\end{figure*}

The NGVS is unique in that it provides photometry for complete samples of 
stellar systems within a single global environment, including galaxies, GCs, 
galactic nuclei, and UCDs.  These systems are often compared to one another 
through their relative luminosities and sizes \citep[\eg][]{Burstein97, MH11, Brodie11}, whereas their relative stellar contents, {\it based on homogeneous 
datasets}, are poorly known.  Given the depth of the NGVS RS, we have a unique 
opportunity to begin filling this gap by examining the colors of faint dwarfs 
and CSS at fixed lumuniosity.

Our samples of GCs, nuclei, and UCDs are drawn from the catalogs of Peng 
\etal~(in preparation), F16, and \cite{Zhang15}, respectively; complete details 
on the selection functions for these samples may be found in those papers.  
Briefly though, GCs and UCDs were both identified via magnitude cuts and the 
\u\i$K$ diagram \citep{Munoz14}, and separated from each other through size 
constraints [$r_h \ge$ 11 pc for UCDs].  The validity of candidate GCs are 
assessed probabilistically and we use only those having a probability $>$ 50\%.  
All UCDs in the \cite{Zhang15} catalog are spectroscopically-confirmed cluster 
members.  Lastly, galactic nuclei were identified by visual inspection of the 
image cutouts for each galaxy and modelled in the 1D surface brightness profiles 
with S{\'e}rsic functions.  For our purposes, we only consider those objects 
classified as unambiguous or possible nuclei in the F16 catalog.

In \Figure{gal-vs-css} we plot the CMDs of galaxies and CSS in Virgo's core 
[left-hand side] and the color distributions for objects with \g~$>$ 18 
[right-hand side]; \u--\g~colors are shown in the upper row and \g--\i~in the 
lower.  Note that we have truncated the CSS samples to 18 $<$ \g~$<$ 22 so that 
our comparisons focus on a common luminosity range.

An obvious difference between the distributions for galaxies and CSS at faint 
luminosities is the latter's extension to very red colors, whereas the former is 
consistent with a single color [\Fig{comp-err}].  This is interesting given that 
CSS have a higher surface density than the faint dwarfs in Virgo's core, 
suggesting that, at fixed luminosity, diffuse systems are forced to be blue 
while concentrated systems can have a wide range of colors.  The nature of red 
CSS is likely explained by a higher metal content, since metallicity more 
strongly affects the colors of quiescent systems than age [see \Fig{comp-k13}].  
Also, the Spearman rank test suggests that nuclei follow CMRs in both 
\u--\g~[$\rho$ = --0.57; $p$ = 4 $\times$ 10$^{-5}$] and \g--\i~[$\rho$ $\sim$ 
--0.5; $p$ = 6 $\times$ 10$^{-4}$], hinting at a possible mass-metallicity 
relation for this population.  A contribution of density to the colors of CSS is 
not obvious though given that many [if not most] of them were produced in the 
vicinity of higher-mass galaxies, and so may owe their enrichment to their local 
environments.  The as-yet uncertain nature of UCDs as either the massive tail of 
the GC population or the bare nuclei of stripped galaxies also raises ambiguity 
on what governs their stellar contents, be it due to internal or external 
factors [\ie self-enrichment versus enriched natal gas clouds].

While it is possible for CSS to be quite red for their luminosities, the 
majority of them have bluer colors, in both \u--\g~and \g--\i, that agree better 
with those of faint RS galaxies.  Closer inspection of the right-half of 
\Fig{gal-vs-css} reveals some tensions between the populations though.  KS tests 
indicate that the null hypothesis of a common parent distribution for galaxies 
and GCs is strongly disfavored for \u--\g~and \g--\i~[$p < 10^{-10}$], whereas 
conclusions vary for UCDs and nuclei depending on the color under consideration 
[$p_{\text{\u--\g}} \sim$ 0.09 and $p_{\text{\g--\i}} < 10^{-4}$ for UCDs; 
$p_{\text{\u--\g}} \sim$ 0.007 and $p_{\text{\g--\i}} \sim$ 0.07 for nuclei].  
The tails in the distributions for the CSS play an important role in these 
tests, but their removal only brings about consistency for the nuclei.  For 
instance, clipping objects with \u--\g $\ge$ 1.2 increases the associated 
$p$--values to 0.18, 0.17, and 0.04 for UCDs, nuclei, and GCs, respectively, 
while $p$ changes to $\sim 10^{-4}$, 0.65, and $< 10^{-4}$ by removing objects 
with \g--\i~$\ge$ 0.85.  We have also fit skewed normal distributions to each 
dataset, finding consistent mean values between galaxies and CSS [except GCs, 
which have a larger value in \g--\i], while the standard deviations for galaxies 
is typically larger than those for CSS.  The evidence for common spectral shapes 
between the majority of CSS and faint galaxies in the core of Virgo is therefore 
conflicting.  An initial assessment of the relative stellar contents within 
these systems, and potential trends with surface density and/or local 
environment, via a joint UV-optical-NIR analysis is desirable to pursue this 
subject further (\eg~Spengler \etal, in preparation).


\section{Comparison to Galaxy Formation Models}\label{sec:comp-model}

\begin{figure*}
 \begin{center}
  \includegraphics[width=1.0\textwidth]{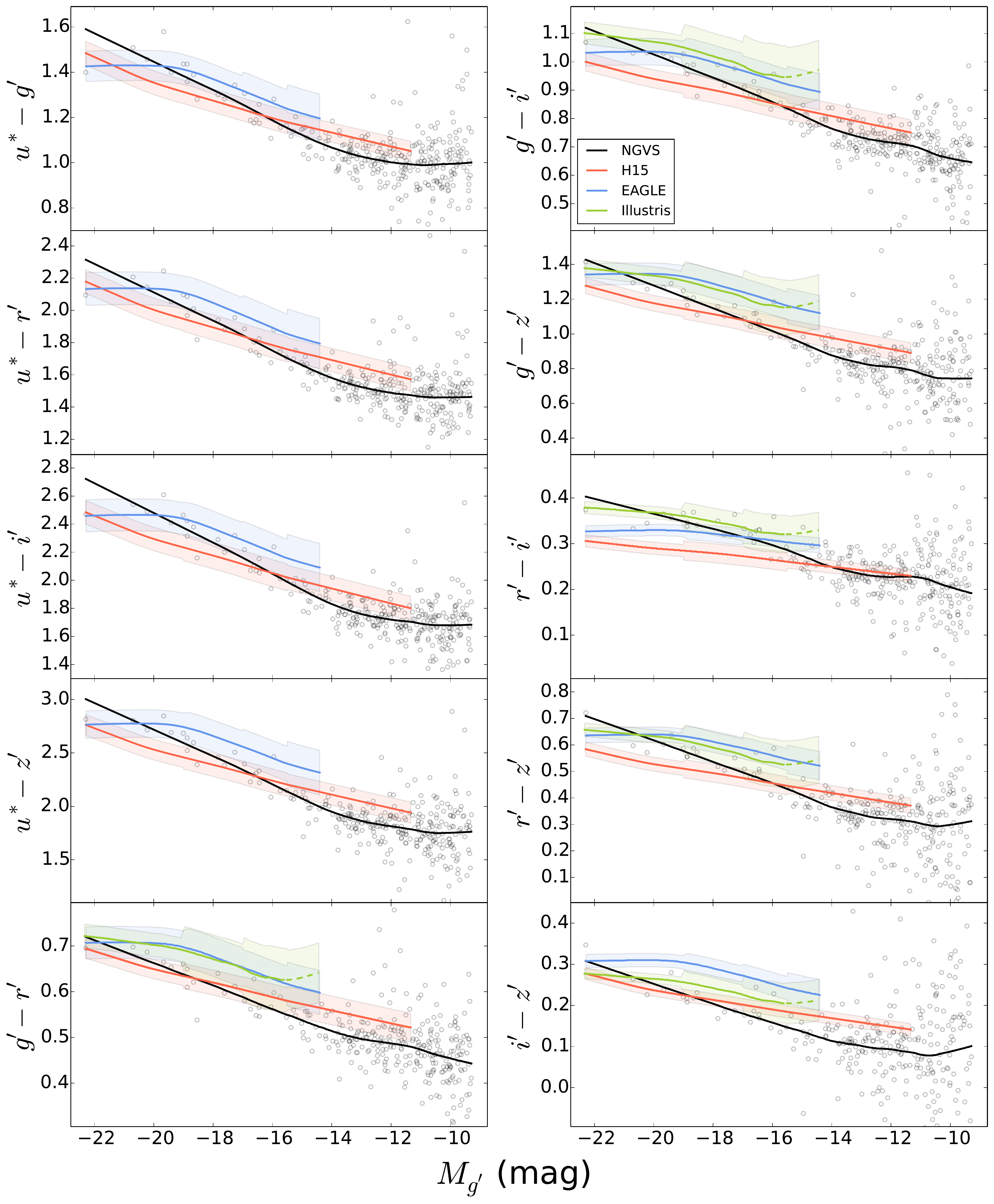}
  \caption{Comparison of the NGVS RS to those from galaxy formation models, with 
  gray circles marking the positions of the observed galaxies.  The shaded 
  region surrounding each model curve indicates the 1-$\sigma$ scatter, measured 
  in five bins of luminosity.  Curves for Illustris do not appear in panels 
  showing \u--band colors since their subhalo catalogs lack those magnitudes.  
  In every color, models uniformly predict a shallower slope for the RS than is 
  observed in cluster cores.}
  \label{fig:comp-mod_rs}
 \end{center}
\end{figure*}

As stated earlier, colors allow us to test our understanding of the star 
formation histories and chemical evolution of galaxies; scaling relations 
therein; and ultimately the physics governing these processes.  Here we explore 
whether current galaxy formation models plausibly explain these subjects by 
reproducing the RS in the core of Virgo.  The main novelty of this comparison 
lies in its focus on the oldest and densest part of a \zz $\sim$ 0 cluster, 
where members have been exposed to extreme external forces, on average, for 
several Gyr \citep{Oman13}.  The nature of our sample dictates that this 
comparison is best suited for galaxies of intermediate-to-low masses, although 
we still include high-mass systems for completeness.  Unless otherwise stated, 
when discussing the slope of the RS, we are referring to the interval --19 
$\lesssim$ \Mg $\lesssim$ --15, where its behavior is more or less linear.

We compare our results to three recent models of galaxy formation: one SAM 
\citep[hereafter H15]{Henriques2015} and two hydrodynamic \citep[Illustris and EAGLE;][]{Vogelsberger14, Schaye15}.  H15 significantly revised the L-Galaxies 
SAM, centered on: (i) increased efficiency of radio-mode AGN feedback; (ii) 
delayed reincoporation of galactic winds [scaling inversely with halo mass]; 
(iii) reduced density threshold for star formation; (iv) AGN heating within 
satellites; and (v) no ram pressure stripping of hot halo gas in low-mass 
groups.  H15 built their model on the Millenium I and II cosmological N-body 
simulations \citep{Springel05, BK09}, enabling them to produce galaxies over a 
mass range of $10^7 <$ \MS $< 10^{12}$ \Msun.  Their revisions helped temper the 
persistent issues of SAMs having too large a blue and red fraction at high and 
low galaxy masses, respectively \citep{Guo11, Henriques13}.

Illustris consists of eight cosmological $N$--body hydro simulations, each 
spanning a volume of $\sim$100$^3$ Mpc$^3$, using the moving-mesh code AREPO.  
This model includes prescriptions for gas cooling; stochastic star formation; 
stellar evolution; gas recycling; chemical enrichment; [kinetic] SNe feedback; 
supermassive black hole [SMBH] seeding, accretion and mergers; and AGN 
feedback.  The simulations differ in terms of the resolution and/or particle 
types/interactions considered; we use the one having the highest resolution and 
a full physics treatment.  EAGLE comprises six simulations with a similar nature 
to Illustris but run with a modified version of the SPH code GADGET 3 instead.  
The simulations differ in terms of resolution, sub-grid physics, or AGN 
parameterization, where the latter variations produce a better match to the \zz 
$\sim$ 0 stellar mass function and high-mass galaxy observables, respectively.
The fiducial model [which we adopt] includes radiative cooling; star formation; 
stellar mass loss; feedback from star formation and AGN; and accretion onto and 
mergers of SMBHs.  Modifications were made to the implementations of stellar 
feedback [formerly kinetic, now thermal], gas accretion by SMBHs [angular 
momentum now included], and the star formation law [metallicity dependence now 
included].  The galaxy populations from Illustris and EAGLE both span a range of 
\MS $\gtrsim 10^{8.5}$ \Msun.

We selected galaxies from the \zz = 0.0 snapshot of H15 that inhabit massive 
halos [$M_h > 10^{14}$ M$_{\odot}$], have non-zero stellar masses, are quenched 
[$sSFR < 10^{-11}$ yr$^{-1}$] and bulge-dominated [$B/T >$ 0.5, by mass]; the 
last constraint aims to weed out highly-reddened spirals.  We query the catalogs 
for both the Millenium I and II simulations, where the latter fills in the 
low-mass end of the galaxy mass function, making this sample of model galaxies 
the best match to the luminosity/mass range of our dataset.  Similar selection 
criteria were used to obtain our samples of Illustris and EAGLE galaxies, except 
that involving $B/T$ since bulge parameters are not included with either 
simulation's catalogs.  We also imposed a resolution cut on Illustris such that 
each galaxy is populated by $\ge$240 star particles [minimum particle mass = 1.5 
$\times$ 10$^4$ \Msun].  A similar cut is implicit in our EAGLE selection as 
SEDs are only available for galaxies having \MS $\gtrsim 10^{8.5}$ \Msun.  
Interestingly, most of the brightest cluster galaxies in EAGLE are not quenched, 
such that we make a second selection to incorporate them in our sample; no such 
issue is found with Illustris.  Broadband magnitudes in the SDSS filters were 
obtained from all three models and transformed to the CFHT/MegaCam system [see 
\Se{comp-prev}].  We note that these magnitudes and the associated colors 
correspond to the {\it total} light of these galaxies.

A final note about this comparison is that we stack clusters from each model 
before analysing its RS.  The high densities of cluster cores make them 
difficult to resolve within cosmological volumes, particularly for hydro 
simulations, leading to small samples for individual clusters.  Stacking is 
therefore needed to enable a meaningful analysis of the model CMD for quenched 
cluster-core galaxies.  H15, Illustris, and EAGLE respectively yield $\sim$15k, 
144, and 157 galaxies lying within 300 kpc of their host halos' centers, which 
is roughly equivalent to the projected size of Virgo's core [as we define it].  
Note that the much larger size of the H15 sample is explained by the greater 
spatial volume it models and the fainter luminosities reached by SAMs [\Mg $\le$ 
--12, compared to \Mg $\lesssim$ --15 for hydro models].

In \Figure{comp-mod_rs} we compare the RS from \Fig{rs} [black] to those from 
H15 [red], Illustris [green], and EAGLE [blue], where the curves for the latter 
were obtained in identical fashion to those for the NGVS.  The shaded regions 
about each model RS convey the 1$\sigma$ scatter within five bins of 
luminosity.  The Illustris RS does not appear in the panels showing \u--band 
colors since their catalogs lack SDSS $u$--band magnitudes.

The clear impression from \Fig{comp-mod_rs} is that no model reproduces the RS 
in Virgo's core, with model slopes being uniformly shallower than observed.  
Two-sample $t-$tests of linear fits to the data and models show that these 
differences are significant at the $P$ = 0.01 level, except for the case of the 
EAGLE models and \g--\r color [$P$ = 0.09].  Further, the H15 RS exhibits no 
sign of the flattening we observe at faint magnitudes; the hydro models 
unfortunately lack the dynamic range needed to evaluate them in this regard.  

The model RSs also differ from one another to varying degrees.  First, H15 
favors a shallower slope than the hydro models.  Second, the color of the H15 RS 
monotonically reddens towards bright magnitudes whereas the hydro RSs turnover 
sharply at \Mg $\lesssim$ --19.  EAGLE and Illustris agree well except for the 
ubiquitos upturn at faint magnitudes in the latter's RS [marked with dashed 
lines].  These upturns are created by the resolution criterion we impose on the 
Illustris catalog and should be disregarded.  Underlying this behavior is the 
fact that lines of constant \MS trace out an approximate anti-correlation in 
color-magnitude space \citep{RC15}, a pattern clearly seen when working with 
richer samples from this model [\eg~galaxies from {\it all} cluster-centric 
radii].  Third, the scatter in H15 is typically the smallest and approximately 
constant with magnitude, whereas those of the hydro models are larger and 
increase towards faint magnitudes, more so for Illustris.  Given that we find 
little intrinsic scatter in the NGVS RS at \Mg $>$ --15 [\Fig{comp-err}], H15 
appears to outperform the hydro models in this regard, although we can only 
trace the latter's scatter to \Mg $\sim$ --15.  Other differences between 
Illustris and EAGLE appear for the colors \g--\i, \r--\i, and \i--\z, in terms 
of turnovers, slopes and/or zeropoints, all of which are significant [$P$ = 
0.01].  It is worth noting that while \Fig{comp-mod_rs} references colors 
measured within 1.0 \Reg~for NGVS galaxies [to maximize their numbers], the 
agreement is not much improved if we use colors from larger apertures.

The conflicting shapes on the RS from data and models could be viewed in one of 
two ways: (i) the core of Virgo is somehow special, or (ii) models fail to 
reproduce the evolution of cluster-core galaxies.  To help demonstrate that the 
latter is more probable, we compare the same models against a separate dataset 
for nearby clusters.  WINGS \citep{Fasano02, Fasano06} is an all-sky survey of a 
complete, X--ray selected sample of 77 galaxy clusters spread over a narrow 
redshift range [\zz = 0.04 -- 0.07].  \cite{Valentinuzzi11} measured the slope 
of the RS for 72 WINGS clusters using $BV$ photometry for galaxies in the range 
--21.5 $\le M_V \le$ --18.  We have done likewise for each well-populated [$N >$ 
100] model cluster, using the \cite{BR07} filter transformations to obtain $BV$ 
photometry from SDSS $gr$--band magnitudes.

\begin{figure}
 \includegraphics[width=0.48\textwidth]{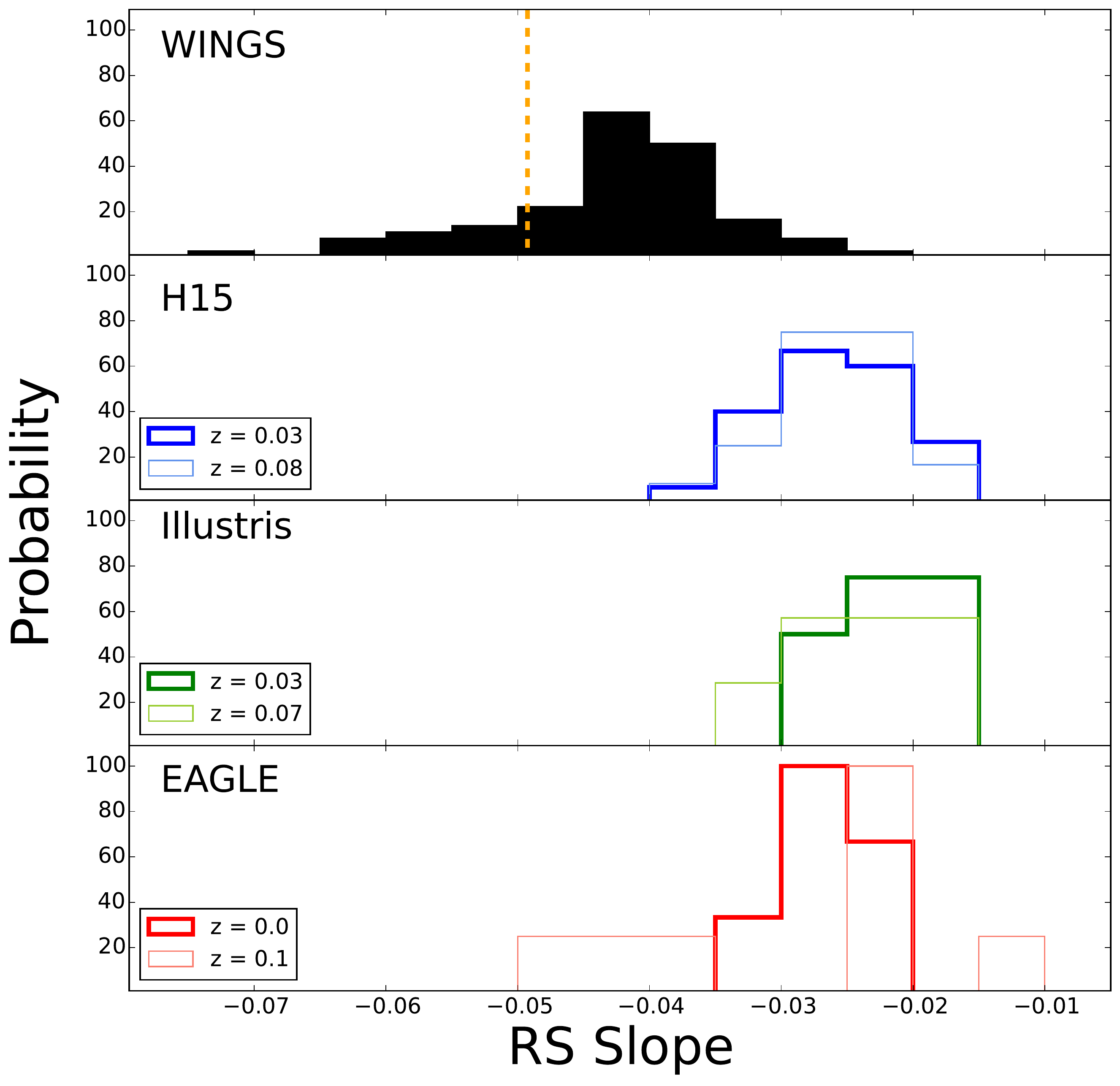}
 \caption{Comparison of RS slopes in real ({\it top panel}) and model clusters 
 ({\it other panels}).  The model slopes are measured from those snapshots which 
 most closely bracket the redshift range of the WINGS clusters [0.03 $\le$ \zz 
 $\le$ 0.07].  In all cases the typical slope within model clusters is shallower 
 than observed.  The dashed line indicates the RS in Virgo's core.}
 \label{fig:comp-mod_slope}
\end{figure}

\Figure{comp-mod_slope} compares the distribution of RS slopes from WINGS and 
galaxy formation models, with the dashed line in the top panel indicating the 
value in Virgo's core, which fits comfortably within the former.  Each model 
distribution is shown for the two closest snapshots to the redshift limits of 
the WINGS sample.  In the case of H15 and Illustris, these snapshots bracket the 
WINGS range quite well, whereas the redshift sampling of EAGLE is notably 
coarser.  The latter fact may be important to explaining the difference between 
the two distributions for this model, since \zz = 0.1 corresponds to a look-back 
time of $\sim$1.3 Gyr.  On the other hand, H15 and Illustris suggest that the RS 
slope does not evolve between \zz = 0.07/0.08 and 0.03.  We have not tried to 
link model clusters across redshifts as parsing merger trees lies beyond the 
scope of this work.  Observations though support the idea of a static slope in 
clusters over the range \zz = 0 -- 1 \citep{Gladders98, Stanford98, Blakeslee03, Ascaso08}.

\Fig{comp-mod_slope} demonstrates that the distributions for the WINGS and model 
clusters are clearly incompatible, with the models, on average, preferring a 
shallower slope for the RS.  The sense of this discrepancy is the same as that 
seen in \Fig{comp-mod_rs} between the core of Virgo and the models.  A caveat 
with the comparisons to WINGS though is that the model slopes have all been 
measured in the respective rest-frames of the clusters.  In other words, the 
model slopes could be biased by differential redshifting of galaxy colors as a 
function of magnitude [\eg~fainter galaxies reddened more than brighter ones].  
To address this, we have simulated the effect of $k-$corrections using the 
median of the EAGLE distribution at \zz = 0.1, finding it would steepen this 
cluster's RS by --0.01.  While significant, we recall that the redshift range 
for the WINGS sample is \zz = 0.04 -- 0.07, such that the mean $k-$correction to 
the model slopes is likely smaller than this value and would therefore not bring 
about better agreement.

Given the value of the above comparisons for testing galaxy formation models, we 
provide in the Appendix parametric fits to the NGVS RS in every color [measured 
at 1 \Reg].  These fits reproduce our LOWESS curves well and enable the wider 
community to perform their own comparisons.


\section{Discussion}\label{sec:disc}

\Figure{rs_demo} indicates that $>$90\% of the galaxy population within the 
innermost $\sim$300 kpc of the Virgo cluster has likely been quenched of star 
formation.  This makes the population ideal for studying the characteristics of 
the RS, such as its shape and intrinsic scatter.  Our analysis demonstrates 
that, in all optical colors, the RS is (a) non-linear and (b) strongly flattens 
in the domain of faint dwarfs.  The former behavior had already been uncovered 
in Virgo, albeit at the bright end (\citealt{Ferrarese06}; JL09), while the 
latter, which is new, begins at --14 $<$ \Mg $<$ --13 [see Appendix], well above 
the completeness limit of the NGVS.  No correlation is observed between color 
and surface brightness, in bins of luminosity, for \Mg $>$ --15, implying that 
the faint-end flattening is not the result of bias or selection effect.

The RS follows the same general shape at \Mg $<$ --14 in each color, which may 
have implications for trends in the stellar populations of these galaxies.  
Assuming that bluer [\eg~\u--\g] and redder [\eg~\g--\z] colors preferentially 
trace mean age and metallicity \citep{Roediger11b}, respectively, the decrease 
in color towards faint magnitudes over the range --19 $\lesssim$ \Mg $\le$ --14 
hints that the populations become younger and less enriched (consistent with 
downsizing; \citealt{Nelan05}), with two exceptions.  The flattening at bright 
magnitudes, seen better in samples that span the full cluster (JL09) and the 
global galaxy population \citep{Baldry04}, signals either a recent burst of star 
formation within these galaxies or an upper limit to galactic chemical 
enrichment.  The latter seems more likely given that the stellar 
mass-metallicity relation for galaxies plateaus at \MS $\gtrsim$ 10$^{11.5}$ 
\Msun~\citep{Gallazzi05}.  The other exception concerns the flattening at the 
faint-end of the RS.

\subsection{What Causes the Faint-End Flattening of the RS?}\label{sec:d-causes}

If colors reasonably trace stellar population parameters [see next sub-section], 
then arguably the most exciting interpretation suggested by the data is that the 
faint dwarfs in Virgo's core have a near-uniform age and metallicity, over a 
range of $\sim$3--4 magnitudes.  This would imply that the known stellar 
population scaling relations for quiescent galaxies of intermediate-to-high mass 
\citep[\eg][]{Choi14} break down at low masses [below $\sim$4 $\times$ 10$^7$ 
\Msun; see Appendix] and, more fundamentally, that the physics governing the 
star formation histories and chemical enrichment of galaxies decouples from mass 
at these scales.

Given the nature of our sample, the above scenario begs the questions of whether 
the faint-end flattening of the RS is caused by the environment, and if so, when 
and where the quenching occurs.  While \cite{Geha12} make the case that dwarfs 
with \MS $<$ 10$^9$ \Msun~must essentially be satellites in order to quench 
(also see \citealt{SB14}; \citealt{Phillips15}; \citealt{Davies16}), we know 
little of the efficiency and timescale of quenching at low satellite masses and 
as a function of host halo mass.  Using Illustris, \cite{Mistani16} showed that, 
on average, the time to quench in low-mass clusters decreases towards low 
satellite masses, from $\sim$5.5 Gyr to $\sim$3 Gyr, over the range 8.5 
$\lesssim$ log \MS $\lesssim$ 10.  \cite{SB14} combine measurements of Local 
Group dwarfs with $N$-body simulations to suggest that, in such groups, galaxies 
of \MS $\lesssim$ 10$^7$ \Msun~quench within 1-2 Gyr of their first pericenter 
passage.  However, \cite{Weisz15} compared HST/WFPC2 star formation histories to 
predicted infall times based on Via Lactea II \citep{Diemand08}, finding that 
many dwarfs in the Local Group likely quenched prior to infall.

In addition to reionization, pre-processing within smaller host halos may play a 
key role in explaining why many Local Group dwarfs ceased forming stars before 
their accretion.  Likewise, pre-processing must also be considered when trying 
to understand issues pertaining to quenching of cluster galaxies \citep[\eg][]{McGee09, DL12, Wetzel13, Hou14, Taranu14}, such as the cause of Virgo's 
flattened RS at faint magnitudes.  \cite{Wetzel13} deduced where satellites of 
\zz = 0 groups/clusters were when they quenched their star formation, by 
modelling SDSS observations of quiescent fractions with mock catalogs.  They 
found that for host halo masses of 10$^{14-15}$ \Msun~the fraction of satellites 
that quenched via pre-processing increases towards lower satellite masses, down 
to their completeness limit of \MS $\sim$ 7 $\times$ 10$^9$ \Msun, largely at 
the expense of quenching in-situ.  Extrapolating this trend to lower satellite 
masses suggests that the majority of the quiescent, low-mass dwarfs in Virgo 
were quenched elsewhere.  This suggestion is consistent with abundance matching 
results for our sample \citep{Grossauer15}, which indicate that only half of the 
core galaxies with \MS = 10$^{6-7}$ \Msun~were accreted by \zz $\sim$ 1 
\citep[see also][]{Oman13}.

\begin{figure}
 \begin{center}
  \includegraphics[width=0.48\textwidth]{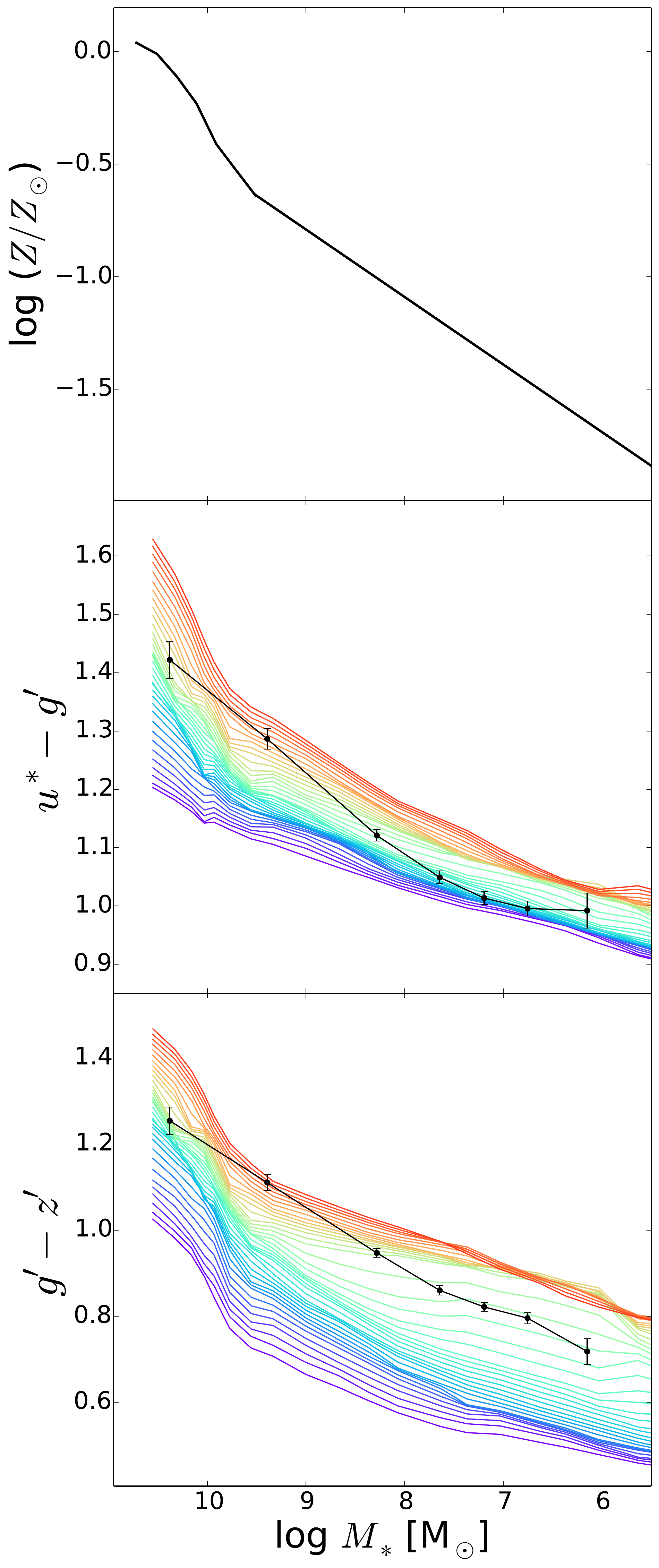}
  \caption{\u--\g~and \g--\z~color-mass relations [middle and bottom panels; 
  black lines] versus those predicted by the FSPS stellar population model 
  [colored lines], constrained by the \cite{Kirby13} mass-metallicity relation 
  [top panel].  Each model relation corresponds to a certain fixed age, ranging 
  between $\sim$2 Gyr [purple] and $\sim$15 Gyr [red] in steps of 0.025 dex.  
  Error bars on the NGVS relations represent standard errors in the mean within 
  bins of luminosity.}
  \label{fig:comp-k13}
 \end{center}
\end{figure}

Assuming that the flattening of the RS reflects an approximate homogeneity in 
stellar contents [\ie constant mean age] and isolated low-mass dwarfs have 
constant star formation histories \citep[\eg][]{Weisz14}, then the low-mass 
dwarfs in Virgo's core must have quenched their star formation coevally.  
Moreover, when coupled with a significant contribution by pre-processing, it is 
implied that these galaxies are highly susceptible to environmental forces, over 
a range of host masses.  This seems plausible given the very high quiescent 
fractions [$>$80\%] for satellites between 10$^6$ $<$ $M_*$/\Msun~$<$ 10$^8$ 
within the Local Volume \citep{Phillips15}, which has led to the idea of a 
threshold satellite mass for effective environmental quenching 
\citep{Geha12, SB14}.

If synchronized quenching of low-mass dwarfs in groups [at least to $\sim$10$^{12}$ \Msun] leads to a flattened faint-end slope of the core RS, we should expect 
to find the same feature for dwarfs drawn from larger cluster-centric radii.  
This follows from the fact that a satellite's cluster-centric radius correlates 
with its infall time \citep{DL12} and that the fraction of satellites accreted 
via groups increases towards low redshift \citep{McGee09}.  Studying the 
properties of the RS as a function of cluster-centric position 
\citep[\eg~see][]{SJ08} will be the focus of a future paper in the NGVS series.

\subsection{Caveats}\label{sec:d-cav}

A major caveat with the above interpretations is that optical colors are not 
unambiguous tracers of population parameters, especially at low metallicities 
\citep{CG10}.  To this point, \cite{Kirby13} have shown that stellar metallicity 
increases monotonically for galaxies from [Fe/H] $\sim$ --2.3 at \MS = 10$^4$ 
\Msun~to slightly super-solar at \MS = 10$^{12}$ \Msun.  Assuming this trend 
holds in all environments, we can check for any conditions under which the RS 
would flatten at faint magnitudes.  In the middle and bottom panels of 
\Figure{comp-k13} we compare the \u--\g~and \g--\z~color-mass relations in 
Virgo's core [black lines] to those predicted by the Flexible Stellar Population 
Synthesis [FSPS] model \citep{Conroy09}, where the Kirby \etal~relation [top 
panel] is used to assign masses to each model metallicity track and lines of 
constant age are colored from purple [$\sim$2 Gyr] to red [$\sim$15 Gyr].  Other 
models \citep[\eg][]{BC03} prove inadequate for our needs due to their coarse 
sampling of metallicity space over the range $Z$ $\sim 4 \times 10^{-4}$ to $4 
\times 10^{-3}$.  Error bars on the NGVS relations reflect standard errors in 
the mean, measured within seven bins of luminosity [having sizes of 0.5-2.0 
dex].  Although we assume single-burst star formation histories for this test, 
qualitatively similar trends are expected for more complex treatments 
(\eg~constant star formation with variable quenching epochs; 
\citealt{Roediger11b}).

Since the intent of \Fig{comp-k13} is to explore an alternative interpretation 
of the faint-end flattening of the RS, we limit our discussion to the range \MS 
$< 10^8$ \Msun, but show the full relations for completeness.  Within that 
range, we find that the data are indeed consistent with Kirby \etal's 
mass-metallicity relation, provided that age does not vary greatly therein.  
Moreover, the color-mass relation for select ages transitions to a flatter slope 
at lower masses.  This confirms our previous statement that it is difficult to 
meaningfully constrain metallicities below a certain level with {\it optical} 
colors [$Z \lesssim 10^{-3}$ in the case of FSPS], even when ages are 
independently known.  The inconsistent ages we would infer from the the 
\u--\g~and \g--\z~colors could likely be ameliorated by lowering the zeropoint 
of the Kirby \etal~relation since the former color responds more strongly to 
metallicity for log($Z/Z_{\odot}) \lesssim$ --1.  The comparisons shown in 
\Fig{comp-k13} therefore cast doubt on whether the flattening of the RS at faint 
magnitudes implies both a constant age {\it and} metallicity for cluster 
galaxies at low masses.  Distinguishing between these scenarios will be more 
rigorously addressed in forthcoming work on the stellar populations of NGVS 
galaxies that incorporates UV and NIR photometry as well.

\subsection{Shortcomings of Galaxy Formation Models}\label{sec:d-short}

Regardless of the uncertainties inherent to the interpretation of optical 
colors, we should expect galaxy formation models to reproduce our observations 
if their physical recipes are correct.  Our test of such models is special in 
that it focuses on the core of a \zz = 0 galaxy cluster, where the 
time-integrated effect of environment on galaxy evolution should be maximal.  
However, \Fig{comp-mod_rs} shows that current models produce a shallower RS than 
observed, in all colors.  This issue is not limited to Virgo's core, as 
\Fig{comp-mod_slope} demonstrates that the distributions of RS slopes for {\it 
entire} model clusters populate shallower values than those measured for other 
nearby clusters.  On a related note, \cite{Licitra16} have shown that clusters 
at \zz $<$ 1 in SAMs suffer from ETG populations with too low an abundance and 
too blue colors, while $\sim$10\% of model clusters have positive RS slopes.  On 
the other hand, \cite{Merson16} found broad consistency between observations and 
SAMs in the zeropoint and slope of the RS in \zz $>$ 1 clusters.  This suggests 
that errors creep into the evolution of cluster galaxies in SAMs at \zz $<$ 1.

The discrepancies indicated here follow upon similar issues highlighted by 
modellers themselves.  H15 showed that their model produces a RS having bluer 
colors than observed in the SDSS for galaxies with \MS $\ge 10^{9.5}$ \Msun.  
\cite{Vogelsberger14} found the Illustris RS suffers the same problem, albeit at 
higher masses [\MS $> 10^{10.5}$ \Msun], while also producing too low of a red 
fraction at \MS $< 10^{11}$ \Msun.  \cite{Trayford15} analyzed the colors of 
EAGLE galaxies, finding that its RS matches that from the GAMA survey 
\citep{Taylor15} for $M_r <$ --20.5, but is too red at fainter magnitudes.  Our 
comparisons build on this work by drawing attention to model treatments of dense 
environments over cosmic time and [hopefully] incentivize modellers to employ 
our dataset in future work, especially as they extend their focus towards lower 
galaxy masses.  To this end, the reader is reminded of the parametric fits to 
the NGVS RS provided in the Appendix.

Naturally, the root of the above discrepancies is tied to errors in the stellar 
populations of model galaxies.  The supplementary material of H15 shows that the 
model {\it exceeds} the mean stellar metallicity of galaxies over the range 
$10^{9.5} <$ \MS $\lesssim 10^{10}$ \Msun~by several tenths of a dex while {\it 
undershooting} measurements at $10^{10.5} <$ \MS $\lesssim 10^{11}$ \Msun~by 
$\sim$0.1--0.2 dex.  The issues with the H15 RS then seems to reflect 
shortcomings in {\it both} the star formation and chemical enrichment histories 
of their model galaxies.  Part of the disagreement facing Illustris likely stems 
from the fact that their galaxies have older stellar populations than observed, 
by as much as 4 Gyr, for \MS $\lesssim 10^{10.5}$ \Msun~\citep{Vogelsberger14}.  
\cite{Schaye15} showed that EAGLE produces a flatter stellar mass-metallicity 
relation than measured from local galaxies due to too much enrichment at \MS 
$\lesssim 10^{10}$ \Msun.  Our inspection of the stellar populations in H15 and 
EAGLE reveals that their cluster-core galaxies, on average, have roughly a 
constant mass-weighted age [$\sim$10-11 Gyr] and follow a shallow 
mass-metallicity relation, with EAGLE metallicities exceeding H15 values by 
$\sim$0.3 dex\footnote{We omit Illustris from this dicussion as their catalogs 
do not provide mean stellar ages of their galaxies.}.  The discrepant colors 
produced by models thus reflect errors in both the star formation histories and 
chemical enrichment of cluster galaxies; for instance, ram pressure stripping 
may be {\it too} effective in quenching cluster dwarfs of star formation 
\citep[\eg][]{Steinhauser16}.

Two critical aspects of the RS that modellers must aim to reproduce are the 
flattenings at both bright and faint magnitudes.  The former is already a 
contentious point between models themselves, with hydro varieties producing a 
turnover and while SAMs continuously increase [\Fig{comp-mod_rs}].  We remind 
the reader that our LOWESS curves are too steep for \Mg $\lesssim$ --19 since 
they essentially represent an extrapolation from intermediate magnitudes; the 
bright-end flattening is clearly visible in other datasets that span the full 
cluster and contain more of such galaxies [\Fig{ngvs-vs-jl09}].  Hydro models 
appear to supercede SAMs in this regard, although it may be argued that their 
turnovers are too sharp.  In the case of EAGLE, however, it is unclear what 
causes this turnover as several of their brightest cluster galaxies are 
star-forming at \zz = 0 while their luminosity-metallicity relation inverts for 
\Mg $\le$ --20.

At present, only SAMs have the requisite depth to check for the flattening seen 
at the faint end of the RS; the effective resolution of cosmological hydro 
models is too low to probe the luminosity function to \Mg $\sim$ --13.  
\Fig{comp-mod_rs} shows that the H15 RS exhibits no obvious change in slope at 
faint magnitudes, let alone the pronounced flattening seen in Virgo.  The 
faint-end flattening is a tantalizing feature of the RS that may hold new 
physical insights into the evolution of cluster galaxies of low mass.  
Addressing the absence of these features should be a focal point for future 
refinements of galaxy formation models.


\section{Conclusions}\label{sec:conc}

We have used homogeneous isophotal photometry in the \u\g\r\i\z~bands for 404 
galaxies belonging to the innermost $\sim$300 kpc of the Virgo cluster to study 
the CMD in a dense environment at \zz = 0, down to stellar masses of $\sim10^6$ 
\Msun.  Our main results are:
\begin{itemize}
 \item The majority of galaxies in Virgo's core populate the RS [red fraction 
 $\sim$ 0.9];
 \item The RS has a non-zero slope at intermediate magnitudes [--19 $<$ \Mg $<$ 
 --14] in all colors, suggesting that stellar age and metallicity both decrease 
 towards lower galaxy masses, and has minimal intrinsic scatter at the faint end;
 \item The RS flattens at both the brightest and faintest magnitudes [\Mg $<$ 
 --19 and \Mg $>$ --14, respectively], where the latter has not been seen before;
 \item Galaxy formation models produce a shallower RS than observed at 
 intermediate magnitudes, for both Virgo and other nearby clusters.  Also, the 
 RS in hydrodynamic models flattens for bright galaxies while that in SAMs 
 varies monotonically over the full range of our dataset.
\end{itemize}
The flattening of the RS at faint magnitudes raises intriguing possibilities 
regarding galaxy evolution and/or cluster formation.  However, these hinge on 
whether the flattening genuinely reflects a homogeneity of stellar populations 
in low-mass galaxies or colors becoming a poor tracer of age/metallicity at low 
metallicities [\eg~log$(Z/Z_{\odot}) \lesssim$ --1.3].  This issue will be 
addressed in a forthcoming paper on the stellar populations of NGVS galaxies.

\begin{figure*}
  \includegraphics[width=1.0\textwidth]{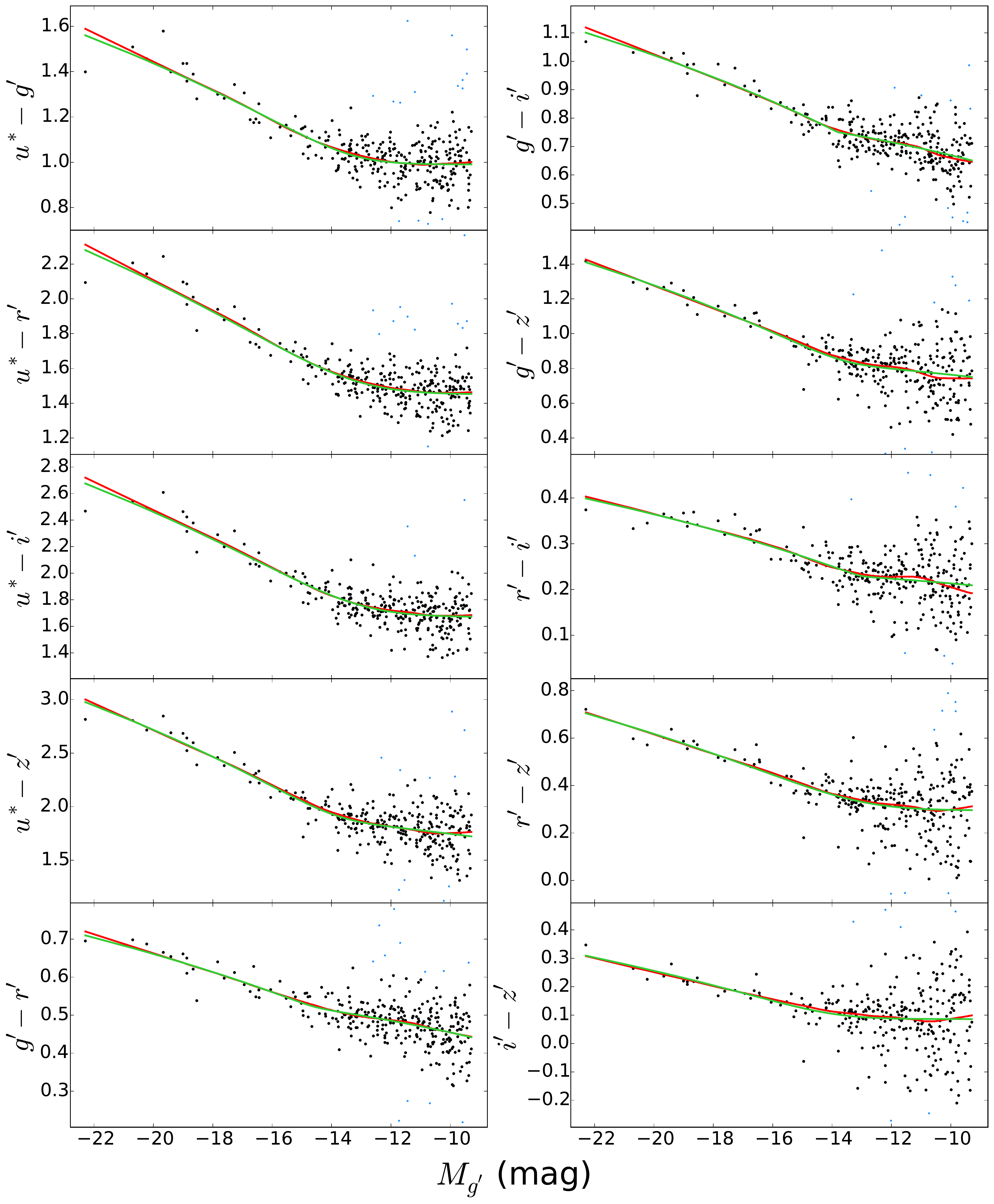}
  \caption{Parameteric fits [green lines] to the RS in Virgo's core, 
  corresponding to the 1.0 \Reg-colors of NGVS galaxies.  These fits are 
  compared to the data themselves [black points] as well as non-parametric 
  [LOWESS] fits.  Points clipped from the each fit are shown in blue.}
  \label{fig:comp_cmr-fit}
\end{figure*}

\bigskip

EWP acknowledges support from the National Natural Science Foundation of China 
under Grant No. 11573002, and from the Strategic Priority Research Program, 
``The Emergence of Cosmological Structures'', of the Chinese Academy of 
Sciences, Grant No. XDB09000105.  C.L. acknowledges the NSFC grants 11203017 and 
11433002.  E.T. acknowledges the NSF grants AST-1010039 and AST-1412504.

This work was supported in part by the Canadian Advanced Network for 
Astronomical Research (CANFAR) which has been made possible by funding from 
CANARIE under the Network-Enabled Platforms program. This research also used the 
facilities of the Canadian Astronomy Data Centre operated by the National 
Research Council of Canada with the support of the Canadian Space Agency.

\appendix\label{sec:app}

\begin{figure*}
  \begin{center}
    \includegraphics[width=0.6\textwidth]{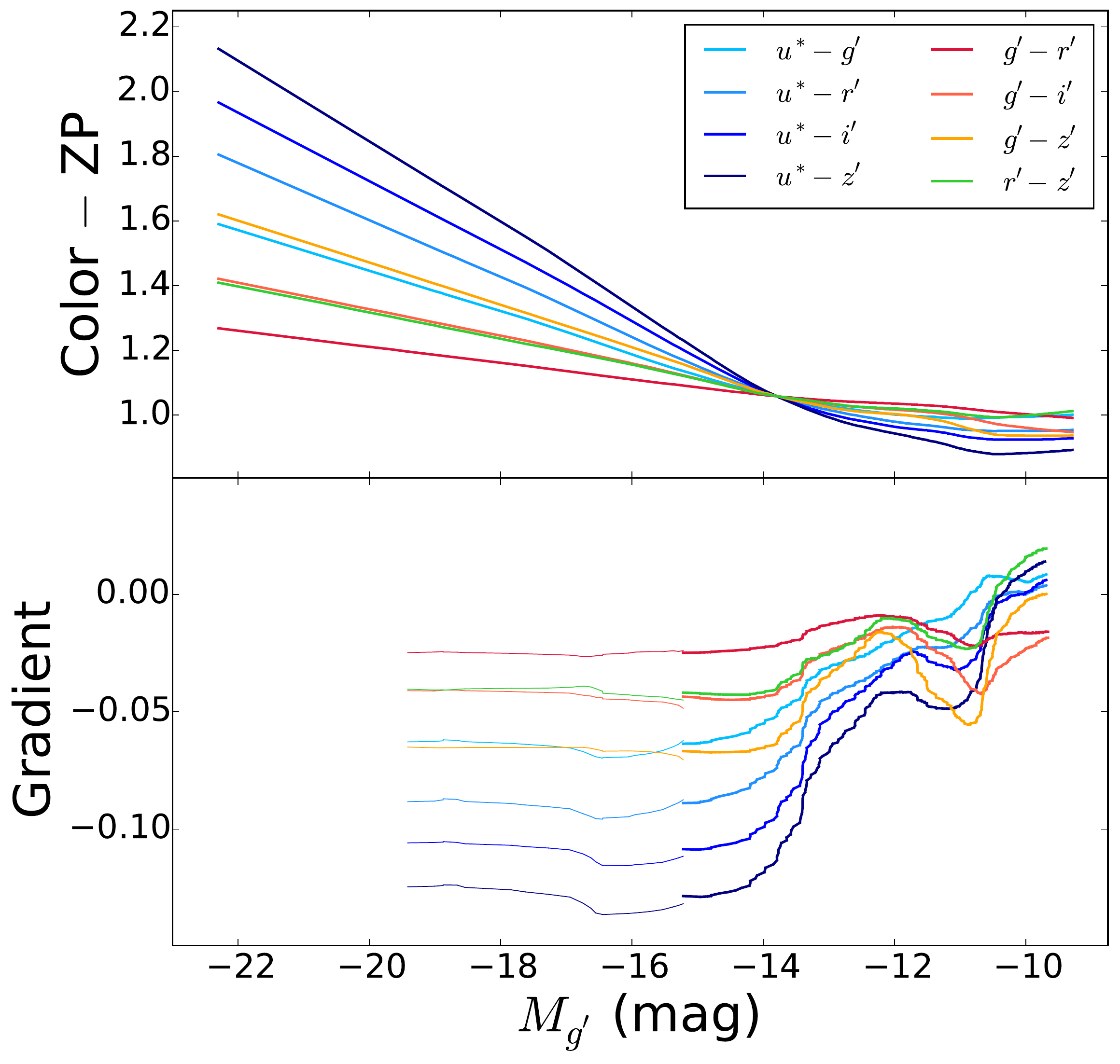}
    \caption{({\it top}) LOWESS fits from \Fig{comp_cmr-fit}, scaled to a common 
    zeropoint at \Mg $\sim$ --14. ({\it bottom}) Local gradient measured along 
    each RS shown in the top panel using a rolling bin of either 9 [thin line] 
    or 51 [thick line] data points; the former bin size allows us to extend our 
    measurements up to bright galaxies.  In all cases, the local gradient begins 
    to flatten in the vicinity of \Mg $\sim$ --15.}
    \label{fig:low-slope}
  \end{center}
\end{figure*}

Here we present parametric fits for the RS in Virgo's core based on the colors 
of our galaxies within 1.0 \Reg.  Our purpose is to enable the wider community, 
particularly modellers, to compare our results to their own through simple 
[continuous] fitting functions.  Motivated by the non-parameteric fits in 
\Fig{rs}, we choose a double power-law to describe the shape of the RS; we 
acknowledge that this choice is made strictly on a phenomenological basis and 
lacks physical motivation.  This function is parameterized as, 

\begin{equation}
\text{Color} = 2^{(\beta_2 - \beta_1)/\alpha}  C_0  \left(\frac{M_{g^{\prime}}}{M_{g^{\prime},0}}\right)^{\beta_2}  \left[1 + \left(\frac{M_{g^{\prime}}}{M_{g^{\prime},0}}\right)^{\alpha}\right]^{(\beta_1 - \beta_2)/\alpha}
\label{eqn:fit}
\end{equation}

where $\beta_1$ and $\beta_2$ represent the asymptotic slopes towards bright and 
faint magnitudes, respectively, while $M_{g^{\prime},0}$ and $C_0$ correspond to 
the magnitude and color of the transition point between the two power-laws, and 
$\alpha$ reflects the sharpness of the transition.

We fit \Eqn{fit} to our data through an iterative non-linear optimization of 
$\chi^2$ following the L-BFGS-B algorithm \citep{Byrd95, Zhu97}, restricting 
$\alpha$, $\beta_1$, and $\beta_2$ to positive values, and $M_{g^{\prime},0}$ 
and $C_0$ to lie in the respective ranges [--20, --8] and [0, 20].  At each 
iteration, $>3\sigma$ outliers are clipped from each CMD; doing so allows the 
fits to better reproduce our LOWESS curves.  We generally achieve convergence 
after 5-6 iterations while the fraction of clipped points is $<$10\% in all 
cases.

Our power-law fits [green curves] are compared to the data [black points] and 
LOWESS fits [red curves] in \Figure{comp_cmr-fit}, while clipped data are 
represented by the blue points.  The best-fit parameters are summarized in 
\Tbl{cmr_fits}, where the final column lists the $rms$ of each fit.  Inspection 
of the $rms$ values and the curves themselves indicates that our parametric fits 
do well in tracing the shape of the RS.

\begin{deluxetable}{lrrrrrr}
 \tabletypesize{100pt}
 \tablecaption{Parameters of double power-law fit to the NGVS RS.}
 \tablehead{
  \colhead{Color} &
  \colhead{$M_{g^{\prime},0}$} &
  \colhead{$C_0$} &
  \colhead{$\beta_1$} &
  \colhead{$\beta_2$} &
  \colhead{$\alpha$} &
  \colhead{$rms$} \\
  \colhead{} &
  \colhead{(mag)} &
  \colhead{(mag)} &
  \colhead{} &
  \colhead{} &
  \colhead{} &
  \colhead{(mag)} \\
  \colhead{(1)} &
  \colhead{(2)} &
  \colhead{(3)} &
  \colhead{(4)} &
  \colhead{(5)} &
  \colhead{(6)} &
  \colhead{(7)}
 }
 \startdata
  \u -- \g & --13.52 & 1.040 & 2.624 & 0.000 & 15.98 & 0.078 \cr
  \u -- \r & --13.62 & 1.552 & 3.871 & 0.000 & 11.51 & 0.091 \cr
  \u -- \i & --13.45 & 1.787 & 4.577 & 0.000 & 11.81 & 0.116 \cr
  \u -- \z & --13.95 & 1.927 & 5.494 & 0.773 & 20.73 & 0.157 \cr
  \g -- \r & --14.57 & 0.522 & 1.036 & 0.392 & 57.14 & 0.047 \cr
  \g -- \i & --13.81 & 0.751 & 1.685 & 0.578 & 1333. & 0.058 \cr
  \g -- \z & --13.74 & 0.852 & 2.808 & 0.413 & 23.39 & 0.124 \cr
  \r -- \i & --13.07 & 0.230 & 0.735 & 0.130 & 96.97 & 0.050 \cr
  \r -- \z & --13.40 & 0.342 & 1.851 & 0.000 & 11.86 & 0.108 \cr
  \i -- \z & --14.15 & 0.107 & 1.133 & 0.000 & 15.92 & 0.102
 \enddata
 \label{tbl:cmr_fits}
\end{deluxetable}

A topic worth exploring with our parametric fits is whether the flattening of 
the RS occurs at a common magnitude for all colors.  This can be done with the 
parameter $M_{g^{\prime},0}$ and \Tbl{cmr_fits} shows that --14 $\le 
M_{g^{\prime},0} \le$ --13, in a large majority of cases.  For \g--\r~and 
\i--\z~the transition magnitude is brighter than --14, which might be explained 
by the fact that these colors sample short wavelength baselines and that the RS 
spans small ranges therein [$\sim$0.25 and 0.15 mag, respectively].  It is also 
likely that the posterior distributions for the parameters in our fit are 
correlated.

Another way to assess the magnitude at which the RS flattens involves measuring 
the local gradient along our LOWESS fits, the advantage being that this approach 
is non-parametric.  \Figure{low-slope} shows our RSs [top panel], scaled to a 
common zeropoint [arbitrarily established at \Mg $\sim$ --14], and the variation 
of the local gradient as a function of magnitude [bottom panel].  We measure the 
local gradient using a running bin of 9 [thin line] or 51 [thick line] data 
points, with the smaller bin allowing us to extend our measurements to brighter 
magnitudes, where our sample is sparse.

The local gradient varies in a consistent way for all colors at \Mg $\le$ --12: 
the gradient is roughly constant and negative at bright magnitudes and becomes 
more positive towards faint magnitudes.  The behaviors of the gradients at \Mg 
$>$ --12 are more irregular as small fluctuations in the LOWESS curves are 
amplified when the gradients hover near zero.  These behaviors are beyond this 
discussion however; we are interested in the locations where the rate of change 
of the gradients is maximized [i.e. the second derivatives of the RSs peak].  
Disregarding the curves at \Mg $>$ --12 then, the bottom panel of 
\Fig{low-slope} shows that the rate of change maximizes in the range --14 $<$ 
\Mg $<$ --13, corresponding to an approximate stellar mass of $\sim4 \times 
10^{7}$ \Msun~\citep{Ferrarese16a}.  The approximate synchronicity of the 
flattening of the RS adds further insight to our main result on the flattening 
of the RS by suggesting a mass scale below which internal processes may cease to 
govern the stellar populations and evolution of dwarf satellites.

\bigskip




\begin{thebibliography}{99}

\bibitem[Adelman-McCarthy et al.(2007)]{AMC07} Adelman-McCarthy, J.~K., Ag{\"u}eros, M.~A., Allam, S.~S., et al.\ 2007, \apjs, 172, 634 
\bibitem[Andreon \& Huertas-Company(2011)]{AHC11} Andreon, S., \& Huertas-Company, M.\ 2011, \aap, 526, A11 
\bibitem[Andreon et al.(2014)]{Andreon14} Andreon, S., Newman, A.~B., Trinchieri, G., et al.\ 2014, \aap, 565, A120 
\bibitem[Ascaso et al.(2008)]{Ascaso08} Ascaso, B., Moles, M., Aguerri, J.~A.~L., S{\'a}nchez-Janssen, R., \& Varela, J.\ 2008, \aap, 487, 453 
\bibitem[Baldry et al.(2004)]{Baldry04} Baldry, I.~K., Glazebrook, K., Brinkmann, J., et al.\ 2004, \apj, 600, 681 
\bibitem[Balogh et al.(2000)]{Balogh00} Balogh, M.~L., Navarro, J.~F., \& Morris, S.~L.\ 2000, \apj, 540, 113 
\bibitem[Balogh et al.(2004)]{Balogh04} Balogh, M.~L., Baldry, I.~K., Nichol, R., et al.\ 2004, \apjl, 615, L101 
\bibitem[Beifiori et al.(2012)]{Beifiori12} Beifiori, A., Courteau, S., Corsini, E.~M., \& Zhu, Y.\ 2012, \mnras, 419, 2497 
\bibitem[Bekki et al.(2003)]{Bekki03} Bekki, K., Couch, W.~J., Drinkwater, M.~J., \& Shioya, Y.\ 2003, \mnras, 344, 399 
\bibitem[Bell et al.(2004)]{Bell04} Bell, E.~F., Wolf, C., Meisenheimer, K., et al.\ 2004, \apj, 608, 752 
\bibitem[Bernardi et al.(2003)]{Bernardi03} Bernardi, M., Sheth, R.~K., Annis, J., et al.\ 2003, \aj, 125, 1866 
\bibitem[Binggeli et al.(1985)]{Binggeli85} Binggeli, B., Sandage, A., \& Tammann, G.~A.\ 1985, \aj, 90, 1681 
\bibitem[Blanton \& Roweis(2007)]{BR07} Blanton, M.~R., \& Roweis, S.\ 2007, \aj, 133, 734 
\bibitem[Blanton \& Moustakas(2009)]{BM09} Blanton, M.~R., \& Moustakas, J.\ 2009, \araa, 47, 159 
\bibitem[Blanton et al.(2003)]{Blanton03} Blanton, M.~R., Hogg, D.~W., Bahcall, N.~A., et al.\ 2003, \apj, 594, 186 
\bibitem[Blakeslee et al.(2003)]{Blakeslee03} Blakeslee, J.~P., Franx, M., Postman, M., et al.\ 2003, \apjl, 596, L143 
\bibitem[Blakeslee et al.(2009)]{Blakeslee09} Blakeslee, J.~P., Jord{\'a}n, A., Mei, S., et al.\ 2009, \apj, 694, 556 
\bibitem[Boselli \& Gavazzi(2014)]{BG14} Boselli, A., \& Gavazzi, G.\ 2014, \aapr, 22, 74 
\bibitem[Boselli et al.(2011)]{Boselli11} Boselli, A., Boissier, S., Heinis, S., et al.\ 2011, \aap, 528, A107 
\bibitem[Boselli et al.(2014)]{Boselli14} Boselli, A., Voyer, E., Boissier, S., et al.\ 2014, \aap, 570, A69 
\bibitem[Boylan-Kolchin et al.(2009)]{BK09} Boylan-Kolchin, M., Springel, V., White, S.~D.~M., Jenkins, A., \& Lemson, G.\ 2009, \mnras, 398, 1150 
\bibitem[Bower et al.(1992)]{Bower92} Bower, R.~G., Lucey, J.~R., \& Ellis, R.~S.\ 1992, \mnras, 254, 601 
\bibitem[Bower et al.(2006)]{Bower06} Bower, R.~G., Benson, A.~J., Malbon, R., et al.\ 2006, \mnras, 370, 645 
\bibitem[Brammer et al.(2011)]{Brammer11} Brammer, G.~B., Whitaker, K.~E., van Dokkum, P.~G., et al.\ 2011, \apj, 739, 24 
\bibitem[Brodie et al.(2011)]{Brodie11} Brodie, J.~P., Romanowsky, A.~J., Strader, J., \& Forbes, D.~A.\ 2011, \aj, 142, 199 
\bibitem[Bruzual \& Charlot(2003)]{BC03} Bruzual, G., \& Charlot, S.\ 2003, \mnras, 344, 1000 
\bibitem[Burstein et al.(1997)]{Burstein97} Burstein, D., Bender, R., Faber, S., \& Nolthenius, R.\ 1997, \aj, 114, 1365 
\bibitem[Byrd et al.(1995)]{Byrd95} Byrd, R.~H., Lu, P., \& Nocedal, J.\ 1995, SIAM Journal on Scientific and Statistical Computing, 16 (5), 1190 
\bibitem[Cassata et al.(2008)]{Cassata08} Cassata, P., Cimatti, A., Kurk, J., et al.\ 2008, \aap, 483, L39 
\bibitem[Cerulo et al.(2016)]{Cerulo16} Cerulo, P., Couch, W.~J., Lidman, C., et al.\ 2016, \mnras, 457, 2209 
\bibitem[Chen et al.(2010)]{Chen10} Chen, C.-W., C{\^o}t{\'e}, P., West, A.~A., Peng, E.~W., \& Ferrarese, L.\ 2010, \apjs, 191, 1 
\bibitem[Choi et al.(2014)]{Choi14} Choi, J., Conroy, C., Moustakas, J., et al.\ 2014, \apj, 792, 95 
\bibitem[Cleveland (1979)]{Cleveland79} Cleveland, W.~S. 1979, Journal of the American Statistical Association, 74 (368), 829 
\bibitem[Conroy \& Gunn(2010)]{CG10} Conroy, C., \& Gunn, J.~E.\ 2010, \apj, 712, 833 
\bibitem[Conroy et al.(2009)]{Conroy09} Conroy, C., Gunn, J.~E., \& White, M.\ 2009, \apj, 699, 486 
\bibitem[Courteau et al.(2007)]{Courteau07} Courteau, S., Dutton, A.~A., van den Bosch, F.~C., et al.\ 2007, \apj, 671, 203 
\bibitem[Crawford et al.(2009)]{Crawford09} Crawford, S.~M., Bershady, M.~A., \& Hoessel, J.~G.\ 2009, \apj, 690, 1158 
\bibitem[Daddi et al.(2007)]{Daddi07} Daddi, E., Dickinson, M., Morrison, G., et al.\ 2007, \apj, 670, 156 
\bibitem[Davies et al.(2016)]{Davies16} Davies, L.~J.~M., Robotham, A.~S.~G., Driver, S.~P., et al.\ 2016, \mnras, 455, 4013 
\bibitem[De Lucia et al.(2007)]{DL07} De Lucia, G., Poggianti, B.~M., Arag{\'o}n-Salamanca, A., et al.\ 2007, \mnras, 374, 809 
\bibitem[De Lucia et al.(2012)]{DL12} De Lucia, G., Weinmann, S., Poggianti, B.~M., Arag{\'o}n-Salamanca, A., \& Zaritsky, D.\ 2012, \mnras, 423, 1277 
\bibitem[De Propris et al.(2013)]{DP13} De Propris, R., Phillipps, S., \& Bremer, M.~N.\ 2013, \mnras, 434, 3469 
\bibitem[de Vaucouleurs(1961)]{dV61} de Vaucouleurs, G.\ 1961, \apjs, 5, 233 
\bibitem[Diemand et al.(2008)]{Diemand08} Diemand, J., Kuhlen, M., Madau, P., et al.\ 2008, \nat, 454, 735 
\bibitem[Drinkwater et al.(2003)]{Drinkwater03} Drinkwater, M.~J., Gregg, M.~D., Hilker, M., et al.\ 2003, \nat, 423, 519 
\bibitem[Driver et al.(2006)]{Driver06} Driver, S.~P., Allen, P.~D., Graham, A.~W., et al.\ 2006, \mnras, 368, 414 
\bibitem[Elbaz et al.(2007)]{Elbaz07} Elbaz, D., Daddi, E., Le Borgne, D., et al.\ 2007, \aap, 468, 33 
\bibitem[Ellis et al.(1997)]{Ellis97} Ellis, R.~S., Smail, I., Dressler, A., et al.\ 1997, \apj, 483, 582 
\bibitem[Faber et al.(2007)]{Faber07} Faber, S.~M., Willmer, C.~N.~A., Wolf, C., et al.\ 2007, \apj, 665, 265 
\bibitem[Fasano et al.(2002)]{Fasano02} Fasano, G., Bettoni, D., D'Onofrio, M., Kj{\ae}rgaard, P., \& Moles, M.\ 2002, \aap, 387, 26 
\bibitem[Fasano et al.(2006)]{Fasano06} Fasano, G., Marmo, C., Varela, J., et al.\ 2006, \aap, 445, 805 
\bibitem[Ferrarese et al.(2006)]{Ferrarese06} Ferrarese, L., C{\^o}t{\'e}, P., Dalla Bont{\`a}, E., et al.\ 2006, \apjl, 644, L21 
\bibitem[Ferrarese et al.(2012)]{Ferrarese12} Ferrarese, L., C{\^o}t{\'e}, P., Cuillandre, J.-C., et al.\ 2012, \apjs, 200, 4 
\bibitem[Ferrarese et al.(2016a)]{Ferrarese16a} Ferrarese, L., C{\^o}t{\'e}, P., S{\'a}nchez-Janssen, R., et al.\ 2016, \apj, 824, 10 
\bibitem[Ferrarese et al.(2016b)]{Ferrarese16b} Ferrarese, L., C{\^o}t{\'e}, P., MacArthur, L.~A., et al.\ 2016, \apj~(accepted) [F16] 
\bibitem[Ferreras et al.(1999)]{Ferreras99} Ferreras, I., Charlot, S., \& Silk, J.\ 1999, \apj, 521, 81 
\bibitem[Fitzpatrick(1999)]{Fitzpatrick99} Fitzpatrick, E.~L.\ 1999, \pasp, 111, 63 
\bibitem[Font et al.(2008)]{Font08} Font, A.~S., Bower, R.~G., McCarthy, I.~G., et al.\ 2008, \mnras, 389, 1619 
\bibitem[Fontana et al.(2009)]{Fontana09} Fontana, A., Santini, P., Grazian, A., et al.\ 2009, \aap, 501, 15 
\bibitem[Gallazzi et al.(2005)]{Gallazzi05} Gallazzi, A., Charlot, S., Brinchmann, J., White, S.~D.~M., \& Tremonti, C.~A.\ 2005, \mnras, 362, 41 
\bibitem[Geha et al.(2012)]{Geha12} Geha, M., Blanton, M.~R., Yan, R., \& Tinker, J.~L.\ 2012, \apj, 757, 85 
\bibitem[Gilbank et al.(2008)]{Gilbank08} Gilbank, D.~G., Yee, H.~K.~C., Ellingson, E., et al.\ 2008, \apj, 673, 742 
\bibitem[Gladders et al.(1998)]{Gladders98} Gladders, M.~D., L{\'o}pez-Cruz, O., Yee, H.~K.~C., \& Kodama, T.\ 1998, \apj, 501, 571 
\bibitem[Grossauer et al.(2015)]{Grossauer15} Grossauer, J., Taylor, J.~E., Ferrarese, L., et al.\ 2015, \apj, 807, 88 
\bibitem[Gobat et al.(2011)]{Gobat11} Gobat, R., Daddi, E., Onodera, M., et al.\ 2011, \aap, 526, A133 
\bibitem[Gonz{\'a}lez et al.(2009)]{Gonzalez09} Gonz{\'a}lez, J.~E., Lacey, C.~G., Baugh, C.~M., Frenk, C.~S., \& Benson, A.~J.\ 2009, \mnras, 397, 1254 
\bibitem[Gonzalez-Perez et al.(2014)]{GP14} Gonzalez-Perez, V., Lacey, C.~G., Baugh, C.~M., et al.\ 2014, \mnras, 439, 264 
\bibitem[Guo et al.(2011)]{Guo11} Guo, Q., White, S., Boylan-Kolchin, M., et al.\ 2011, \mnras, 413, 101 
\bibitem[Haines et al.(2015)]{Haines15} Haines, C.~P., Pereira, M.~J., Smith, G.~P., et al.\ 2015, \apj, 806, 101 
\bibitem[Hao et al.(2009)]{Hao09} Hao, J., Koester, B.~P., Mckay, T.~A., et al.\ 2009, \apj, 702, 745 
\bibitem[Harris(2016)]{Harris16} Harris, W.~E.\ 2016, \aj, 151, 102 
\bibitem[Henriques et al.(2013)]{Henriques13} Henriques, B.~M.~B., White, S.~D.~M., Thomas, P.~A., et al.\ 2013, \mnras, 431, 3373 
\bibitem[Henriques et al.(2015)]{Henriques2015} Henriques, B.~M.~B., White, S.~D.~M., Thomas, P.~A., et al.\ 2015, \mnras, 451, 2663 
\bibitem[Hilton et al.(2009)]{Hilton09} Hilton, M., Stanford, S.~A., Stott, J.~P., et al.\ 2009, \apj, 697, 436 
\bibitem[Holden et al.(2004)]{Holden04} Holden, B.~P., Stanford, S.~A., Eisenhardt, P., \& Dickinson, M.\ 2004, \aj, 127, 2484 
\bibitem[Hou et al.(2014)]{Hou14} Hou, A., Parker, L.~C., \& Harris, W.~E.\ 2014, \mnras, 442, 406 
\bibitem[Ilbert et al.(2010)]{Ilbert10} Ilbert, O., Salvato, M., Le Floc'h, E., et al.\ 2010, \apj, 709, 644 
\bibitem[Jaff{\'e} et al.(2011)]{Jaffe11} Jaff{\'e}, Y.~L., Arag{\'o}n-Salamanca, A., De Lucia, G., et al.\ 2011, \mnras, 410, 280 
\bibitem[Janz \& Lisker(2009)]{JL09} Janz, J., \& Lisker, T.\ 2009, \apjl, 696, L102 [JL09] 
\bibitem[Johnson et al.(2013)]{Johnson13} Johnson, B.~D., Weisz, D.~R., Dalcanton, J.~J., et al.\ 2013, \apj, 772, 8 
\bibitem[Kennicutt \& Evans(2012)]{KE12} Kennicutt, R.~C., \& Evans, N.~J.\ 2012, \araa, 50, 531 
\bibitem[Kim et al.(2010)]{Kim10} Kim, S., Rey, S.-C., Lisker, T., \& Sohn, S.~T.\ 2010, \apjl, 721, L72 
\bibitem[Kirby et al.(2013)]{Kirby13} Kirby, E.~N., Cohen, J.~G., Guhathakurta, P., et al.\ 2013, \apj, 779, 102 
\bibitem[Kodama \& Arimoto(1997)]{KA97} Kodama, T., \& Arimoto, N.\ 1997, \aap, 320, 41 
\bibitem[Kormendy \& Ho(2013)]{KH13} Kormendy, J., \& Ho, L.~C.\ 2013, \araa, 51, 511 
\bibitem[Kriek et al.(2008)]{Kriek08} Kriek, M., van der Wel, A., van Dokkum, P.~G., Franx, M., \& Illingworth, G.~D.\ 2008, \apj, 682, 896 
\bibitem[Lelli et al.(2016)]{Lelli16} Lelli, F., McGaugh, S.~S., \& Schombert, J.~M.\ 2016, \apjl, 816, L14 
\bibitem[Licitra et al.(2016)]{Licitra16} Licitra, R., Mei, S., Raichoor, A., Erben, T., \& Hildebrandt, H.\ 2016, \mnras, 455, 3020 
\bibitem[Lidman et al.(2008)]{Lidman08} Lidman, C., Rosati, P., Tanaka, M., et al.\ 2008, \aap, 489, 981 
\bibitem[Liu et al.(2015)]{Liu15} Liu, C., Peng, E.~W., C{\^o}t{\'e}, P., et al.\ 2015, \apj, 812, 34 
\bibitem[Lu et al.(2009)]{Lu09} Lu, T., Gilbank, D.~G., Balogh, M.~L., \& Bognat, A.\ 2009, \mnras, 399, 1858 
\bibitem[Marchesini et al.(2014)]{Marchesini14} Marchesini, D., Muzzin, A., Stefanon, M., et al.\ 2014, \apj, 794, 65 
\bibitem[McCall(2004)]{McCall04} McCall, M.~L.\ 2004, \aj, 128, 2144 
\bibitem[McGaugh(2012)]{McGaugh12} McGaugh, S.~S.\ 2012, \aj, 143, 40 
\bibitem[McGee et al.(2009)]{McGee09} McGee, S.~L., Balogh, M.~L., Bower, R.~G., Font, A.~S., \& McCarthy, I.~G.\ 2009, \mnras, 400, 937 
\bibitem[Mei et al.(2006)]{Mei06} Mei, S., Holden, B.~P., Blakeslee, J.~P., et al.\ 2006, \apj, 644, 759 
\bibitem[Mei et al.(2007)]{Mei07} Mei, S., et al.\ 2007, ApJ, 655, 144 
\bibitem[Mei et al.(2009)]{Mei09} Mei, S., Holden, B.~P., Blakeslee, J.~P., et al.\ 2009, \apj, 690, 42 
\bibitem[Menci et al.(2008)]{Menci08} Menci, N., Rosati, P., Gobat, R., et al.\ 2008, \apj, 685, 863-874 
\bibitem[Merson et al.(2016)]{Merson16} Merson, A.~I., Baugh, C.~M., Gonzalez-Perez, V., et al.\ 2016, \mnras, 456, 1681 
\bibitem[Misgeld \& Hilker(2011)]{MH11} Misgeld, I., \& Hilker, M.\ 2011, \mnras, 414, 3699 
\bibitem[Mistani et al.(2016)]{Mistani16} Mistani, P.~A., Sales, L.~V., Pillepich, A., et al.\ 2016, \mnras, 455, 2323 
\bibitem[Mu{\~n}oz et al.(2014)]{Munoz14} Mu{\~n}oz, R.~P., Puzia, T.~H., Lan{\c c}on, A., et al.\ 2014, \apjs, 210, 4 
\bibitem[Muzzin et al.(2009)]{Muzzin09} Muzzin, A., Wilson, G., Yee, H.~K.~C., et al.\ 2009, \apj, 698, 1934 
\bibitem[Muzzin et al.(2013)]{Muzzin13} Muzzin, A., Marchesini, D., Stefanon, M., et al.\ 2013, \apj, 777, 18 
\bibitem[Nelan et al.(2005)]{Nelan05} Nelan, J.~E., Smith, R.~J., Hudson, M.~J., et al.\ 2005, \apj, 632, 137 
\bibitem[Noeske et al.(2007)]{Noeske07} Noeske, K.~G., Weiner, B.~J., Faber, S.~M., et al.\ 2007, \apjl, 660, L43 
\bibitem[Oman et al.(2013)]{Oman13} Oman, K.~A., Hudson, M.~J., \& Behroozi, P.~S.\ 2013, \mnras, 431, 2307 
\bibitem[Papovich et al.(2010)]{Papovich10} Papovich, C., Momcheva, I., Willmer, C.~N.~A., et al.\ 2010, \apj, 716, 1503 
\bibitem[Peng et al.(2002)]{Peng02} Peng, C.~Y., Ho, L.~C., Impey, C.~D., \& Rix, H.-W.\ 2002, \aj, 124, 266 
\bibitem[Pozzetti et al.(2010)]{Pozzetti10} Pozzetti, L., Bolzonella, M., Zucca, E., et al.\ 2010, \aap, 523, A13 
\bibitem[Pfeffer \& Baumgardt(2013)]{PB13} Pfeffer, J., \& Baumgardt, H.\ 2013, \mnras, 433, 1997 
\bibitem[Phillips et al.(2015)]{Phillips15} Phillips, J.~I., Wheeler, C., Cooper, M.~C., et al.\ 2015, \mnras, 447, 698 
\bibitem[Poggianti et al.(2001)]{Poggianti01} Poggianti, B.~M., Bridges, T.~J., Mobasher, B., et al.\ 2001, \apj, 562, 689 
\bibitem[Roediger \& Courteau(2015)]{RC15} Roediger, J.~C., \& Courteau, S.\ 2015, \mnras, 452, 3209 
\bibitem[Roediger et al.(2011a)]{Roediger11a} Roediger, J.~C., Courteau, S., McDonald, M., \& MacArthur, L.~A.\ 2011, \mnras, 416, 1983 
\bibitem[Roediger et al.(2011b)]{Roediger11b} Roediger, J.~C., Courteau, S., MacArthur, L.~A., \& McDonald, M.\ 2011, \mnras, 416, 1996 
\bibitem[Romeo et al.(2008)]{Romeo08} Romeo, A.~D., Napolitano, N.~R., Covone, G., et al.\ 2008, \mnras, 389, 13 
\bibitem[Romeo et al.(2015)]{Romeo15} Romeo, A.~D., Kang, X., Contini, E., et al.\ 2015, \aap, 581, A50 
\bibitem[Rudnick et al.(2009)]{Rudnick09} Rudnick, G., von der Linden, A., Pell{\'o}, R., et al.\ 2009, \apj, 700, 1559 
\bibitem[S{\'a}nchez-Janssen et al.(2008)]{SJ08} S{\'a}nchez-Janssen, R., Aguerri, J.~A.~L., \& Mu{\~n}oz-Tu{\~n}{\'o}n, C.\ 2008, \apjl, 679, L77 
\bibitem[S{\'a}nchez-Janssen et al.(2016)]{SJ16} S{\'a}nchez-Janssen, R., Ferrarese, L., MacArthur, L.~A., et al.\ 2016, \apj, 820, 69 
\bibitem[Schaye et al.(2015)]{Schaye15} Schaye, J., Crain, R.~A., Bower, R.~G., et al.\ 2015, \mnras, 446, 521 
\bibitem[Schawinski et al.(2014)]{Schawinski14} Schawinski, K., Urry, C.~M., Simmons, B.~D., et al.\ 2014, \mnras, 440, 889 
\bibitem[Schlegel et al.(1998)]{Schlegel98} Schlegel, D.~J., Finkbeiner, D.~P., \& Davis, M.\ 1998, \apj, 500, 525
\bibitem[Seth et al.(2014)]{Seth14} Seth, A.~C., van den Bosch, R., Mieske, S., et al.\ 2014, \nat, 513, 398 
\bibitem[Slater \& Bell(2014)]{SB14} Slater, C.~T., \& Bell, E.~F.\ 2014, \apj, 792, 141 
\bibitem[Smail et al.(1998)]{Smail98} Smail, I., Edge, A.~C., Ellis, R.~S., \& Blandford, R.~D.\ 1998, \mnras, 293, 124 
\bibitem[Spitler et al.(2012)]{Spitler12} Spitler, L.~R., Labb{\'e}, I., Glazebrook, K., et al.\ 2012, \apjl, 748, L21 
\bibitem[Springel et al.(2005)]{Springel05} Springel, V., White, S.~D.~M., Jenkins, A., et al.\ 2005, \nat, 435, 629 
\bibitem[Stanford et al.(1998)]{Stanford98} Stanford, S.~A., Eisenhardt, P.~R., \& Dickinson, M.\ 1998, \apj, 492, 461 
\bibitem[Stanford et al.(2012)]{Stanford12} Stanford, S.~A., Brodwin, M., Gonzalez, A.~H., et al.\ 2012, \apj, 753, 164 
\bibitem[Steinhauser et al.(2016)]{Steinhauser16} Steinhauser, D., Schindler, S., \& Springel, V.\ 2016, \aap, 591, A51 
\bibitem[Stott et al.(2007)]{Stott07} Stott, J.~P., Smail, I., Edge, A.~C., et al.\ 2007, \apj, 661, 95 
\bibitem[Strateva et al.(2001)]{Strateva01} Strateva, I., Ivezi{\'c}, {\v Z}., Knapp, G.~R., et al.\ 2001, \aj, 122, 1861 
\bibitem[Strazzullo et al.(2013)]{Strazzullo13} Strazzullo, V., Gobat, R., Daddi, E., et al.\ 2013, \apj, 772, 118 
\bibitem[Tanaka et al.(2005)]{Tanaka05} Tanaka, M., Kodama, T., Arimoto, N., et al.\ 2005, \mnras, 362, 268 
\bibitem[Taranu et al.(2014)]{Taranu14} Taranu, D.~S., Hudson, M.~J., Balogh, M.~L., et al.\ 2014, \mnras, 440, 1934 
\bibitem[Taylor et al.(2015)]{Taylor15} Taylor, E.~N., Hopkins, A.~M., Baldry, I.~K., et al.\ 2015, \mnras, 446, 2144 
\bibitem[Terlevich et al.(1999)]{Terlevich99} Terlevich, A.~I., Kuntschner, H., Bower, R.~G., Caldwell, N., \& Sharples, R.~M.\ 1999, \mnras, 310, 445 
\bibitem[Thomas et al.(2005)]{Thomas05} Thomas, D., Maraston, C., Bender, R., \& Mendes de Oliveira, C.\ 2005, \apj, 621, 673 
\bibitem[Trayford et al.(2015)]{Trayford15} Trayford, J.~W., Theuns, T., Bower, R.~G., et al.\ 2015, \mnras, 452, 2879 
\bibitem[Valentinuzzi et al.(2011)]{Valentinuzzi11} Valentinuzzi, T., Poggianti, B.~M., Fasano, G., et al.\ 2011, \aap, 536, A34 
\bibitem[Visvanathan \& Sandage(1977)]{VS77} Visvanathan, N., \& Sandage, A.\ 1977, \apj, 216, 214 
\bibitem[Vogelsberger et al.(2014)]{Vogelsberger14} Vogelsberger, M., Genel, S., Springel, V., et al.\ 2014, \mnras, 444, 1518 
\bibitem[Wehner \& Harris(2006)]{WH06} Wehner, E.~H., \& Harris, W.~E.\ 2006, \apjl, 644, L17 
\bibitem[Weinmann et al.(2011)]{Weinmann11} Weinmann, S.~M., Lisker, T., Guo, Q., Meyer, H.~T., \& Janz, J.\ 2011, \mnras, 416, 1197 
\bibitem[Weisz et al.(2014)]{Weisz14} Weisz, D.~R., Dolphin, A.~E., Skillman, E.~D., et al.\ 2014, \apj, 789, 147 
\bibitem[Weisz et al.(2015)]{Weisz15} Weisz, D.~R., Dolphin, A.~E., Skillman, E.~D., et al.\ 2015, \apj, 804, 136 
\bibitem[Wetzel et al.(2013)]{Wetzel13} Wetzel, A.~R., Tinker, J.~L., Conroy, C., \& van den Bosch, F.~C.\ 2013, \mnras, 432, 336 
\bibitem[Willmer et al.(2006)]{Willmer06} Willmer, C.~N.~A., Faber, S.~M., Koo, D.~C., et al.\ 2006, \apj, 647, 853 
\bibitem[Wilson et al.(2009)]{Wilson09} Wilson, G., Muzzin, A., Yee, H.~K.~C., et al.\ 2009, \apj, 698, 1943 
\bibitem[York et al.(2000)]{York00} York, D.~G., Adelman, J., Anderson, J.~E., Jr., et al.\ 2000, \aj, 120, 1579 
\bibitem[Zaritsky et al.(2012)]{Zaritsky12} Zaritsky, D., Zabludoff, A.~I., \& Gonzalez, A.~H.\ 2012, \apj, 748, 15 
\bibitem[Zhang et al.(2015)]{Zhang15} Zhang, H.-X., Peng, E.~W., C{\^o}t{\'e}, P., et al.\ 2015, \apj, 802, 30 
\bibitem[Zhu et al.(1997)]{Zhu97} Zhu, C., Byrd, R.~H., \& Nocedal, J.\ 1997, ACM Transactions on Mathematical Software, 23 (4), 550 
\bibitem[Zhu et al.(2014)]{Zhu14} Zhu, L., Long, R.~J., Mao, S., et al.\ 2014, \apj, 792, 59 

\end{thebibliography}
\end{document}